\documentclass[aps,pre,twocolumn,longbibliography]{revtex4-2}

\usepackage{graphicx}
\usepackage{color}
\usepackage{graphicx}
\usepackage{amsmath}
\usepackage{dcolumn}

\usepackage{epsfig}
\usepackage{graphics}
\usepackage{latexsym}
\usepackage{amsfonts}
\usepackage{amssymb}

\usepackage{bbm}
\usepackage{bbold}
\usepackage{mathtools}
\usepackage{booktabs}
\usepackage{placeins} 
\usepackage{stmaryrd} 

\DeclareMathOperator{\Li}{Li}
\DeclareMathOperator{\sech}{sech}
\DeclareMathOperator{\sign}{sign}

\newcommand{\al}{\alpha_\mathrm{o}}
\newcommand{\alb}{\bar{\alpha}_\mathrm{o}}
\newcommand{\arod}{a_\mathrm{E}} 
\newcommand{\as}{a_\mathrm{s}}
\newcommand{\avg} [1]{\langle #1 \rangle}
\newcommand{\bet}{\beta}

\newcommand{\bcontaminans}{{\it B.~contaminans}}
\newcommand{\bh}{\hat{b}}
\newcommand{\Bc}{B_\mathrm{o}}
\newcommand{\Bchat}{B_\mathrm{c}}
\newcommand{\brod}{b_\mathrm{E}} 
\newcommand{\capti}[1]{{\bf #1}}
\newcommand{\catalan}{\mathcal{C}}
\newcommand{\consB}{C_\mathcal{B}}
\newcommand{\cin}{c_\mathrm{in}}
\newcommand{\cop}{{c_\mathrm{o}'}}
\newcommand{\costheta}{\cos \theta}
\newcommand{\cout}{c_\mathrm{out}}
\newcommand{\csat}{c_\mathrm{s}}
\newcommand{\cs}{c^*}
\newcommand{\dd}{\mathrm{d}}
\newcommand{\dif}{D}
\newcommand{\difb}{\mathcal{D}}
\newcommand{\difrot}{D_\mathrm{r}}
\newcommand{\eps}{\epsilon}
\newcommand{\eqdef}{\equiv}
\newcommand{\etal}{{\it et al.}}
\newcommand{\fa}{f_\mathrm{a}}
\newcommand{\fm}{f_\mathrm{m}}

\newcommand{\fmid}{\breve{f}} 
\newcommand{\gam}{\gamma}
\newcommand{\gami}{\gam^{-1}}
\newcommand{\Gamo}{\Gamma_\mathrm{o}}

\newcommand{\Gapp}{G_\mathrm{app}}
\newcommand{\intI}{\mathcal{I}} 
\newcommand{\intpi}{\int_{-\pi}^\pi}
\newcommand{\inv}{$^{-1}$}
\newcommand{\kap}{\kappa}
\newcommand{\kB}{k_\mathrm{B}}
\newcommand{\ko}{k}
\newcommand{\kom}{\ko}
\newcommand{\krn}{\mathsf{k}}

\newcommand{\krnetc}{\krn(c, c')}
\newcommand{\Krn}{\mathsf{K}}
\newcommand{\la}{l_\mathrm{a}}    
\newcommand{\lam}{\lambda}                   
\newcommand{\lamhy}{\Lambda} 
\newcommand{\lamlr}{\lam_{LR}}
\newcommand{\lamrl}{\lam_{RL}}
\newcommand{\lamo}{\lam_\mathrm{o}}
\newcommand{\lamoi}{\lamo^{-1}}
\newcommand{\len}{l}

\newcommand{\lm}{l_-}
\newcommand{\leno}{\len_\mathrm{o}}
\newcommand{\lp}{l_+}
 
\newcommand{\luni}{l_\mathrm{u}} 
\newcommand{\Lc}{L}

\newcommand{\Mgryphiwaldensefull}{{\it Magnetospirillum gryphiswaldense}}
\newcommand{\Mgryphiwaldense}{{\it M.~gryphiswaldense}}
\newcommand{\microm}{µm}
\newcommand{\micromsi}{\microm$\,$\si}
\newcommand{\mo}{m_\mathrm{o}}

\newcommand{\nor}{\mathcal{N}}

\newcommand{\Ntotlat}{\mathcal{N}}
\newcommand{\nupar}{\nu} 
\newcommand{\ot}{\,}
\newcommand{\opT}{\mathrm{T}}
\newcommand{\pA}{\mathcal{A}}
\newcommand{\pB}{\mathcal{B}}
\newcommand{\pC}{\mathcal{C}}

\newcommand{\piso}{p_\mathrm{iso}}
\newcommand{\Phicons}{\Phi}  

\newcommand{\pres}{\mathcal{P}}
\newcommand{\Psim}{\Psi_{-}} 
\newcommand{\Psip}{\Psi_{+}} 
\newcommand{\pv}{\mathrm{p.v.}}
\newcommand{\qo}{q_\mathrm{o}}
\newcommand{\Res}{\mathcal{R}}
\newcommand{\rhom}{\rho}
\newcommand{\rondF}{\mathcal{F}} 
\newcommand{\si}{s\inv}

\newcommand{\sigo}{\sigma_\mathrm{o}}
\newcommand{\tauo}{\tau_\mathrm{o}}
\newcommand{\tensE}{\mathbf{E}} 
\newcommand{\tensI}{\mathbf{I}} 
\newcommand{\tensT}{\mathbf{T}}
\newcommand{\tensTa}{\tensT_\mathrm{a}}
\newcommand{\tensTBr}{\tensT_\mathrm{Br}}
\newcommand{\tensTm}{\tensT_\mathrm{m}}
\newcommand{\tensTv}{\tensT_\mathrm{v}}
\newcommand{\tensW}{\mathbf{W}} 
\newcommand{\thetaturn}{\theta_\mathrm{turn}}
\newcommand{\va}{v_\mathrm{a}} 
\newcommand{\vfa}{\vc{f}_\mathrm{a}} 
\newcommand{\vfm}{\vc{f}_\mathrm{m}} 
\newcommand{\vint}{v^\mathrm{int}}
\newcommand{\vm}{v_\mathrm{m}}
\newcommand{\vmo}{\vm^0}
\newcommand{\Vscale}{V}
\newcommand{\vo}{v_\mathrm{o}}
\newcommand{\volfrac}{\Phi}
\newcommand{\xione}{\xi_{1}}
\newcommand{\xil}{\zeta} 
\newcommand{\xilp}{\hat{\xil}}
\newcommand{\xm}{x_\mathrm{m}}
\newcommand{\xs}{x^*}
\newcommand{\xt}{\tilde{x}}

\newcommand{\vc}[1]{\mathbf{#1}}
\newcommand{\vB}{\vc{B}}
\newcommand{\vcr}{\vc{r}}
\newcommand{\ve}{\vc{e}}
\newcommand{\veB}{\ve_\mathrm{B}}

\newcommand{\vey}{\vc{e}_y}

\newcommand{\vv}{\vc{v}}

\usepackage{color}
\usepackage[dvipsnames]{xcolor}

\usepackage[hidelinks]{hyperref}
\usepackage{comment}
\hypersetup{urlcolor=blue}

\begin{document}

\title{Exact model of aerotactic band: \\
From Fokker-Planck equation to  band structure and fluid flow} 

\author{F.~Detcheverry}  
\affiliation{University of Lyon, Universit\'{e} Claude Bernard Lyon 1, CNRS, Institut Lumi\`{e}re Mati\`{e}re, F-69622, Villeurbanne, France}

\begin{abstract}
A variety of bacterial species spontaneously assemble in aerotactic band,  
local accumulation at a fixed distance from the air-water interface. 
Although the phenomenon is long known,  
its modelling is so far limited to  mesoscopic, one-dimensional or numerical descriptions.    
We investigate band properties  at the microscopic scale 
using exact  solutions to the Fokker-Planck equation.  
First, 
we  show that the interplay between oxygen consumption and  tumbling modulation  
is governed by a third-order nonlinear differential equation  
relating the oxygen concentration to the aerotactic response. 
For two model aerotactic behaviors, 
we present analytical solutions and discuss the resulting  band structure. 
Second, 
we investigate how an aerotactic band of magnetotactic bacteria in a magnetic field 
induces a spontaneous  fluid flow, 
as observed in experiments [Marmol \etal, {\it arXiv} 2025]. 
In the low field limit,
we determine the bacterial distribution  and the active stress tensor. 
Using the Green function of the hydrodynamic problem, 
we obtain a prediction for the fluid flow 
that is both simple and consistent with observations.  
Altogether, our results provide a model system of aerotactic band 
and solid ground to analyze aerotaxis-driven self-organization.  
\end{abstract}

\date{\today}

\maketitle

\section{Introduction}

Oxygen is a shaping factor for living matter, 
a fact that may date back to the early chapters of life 
since  the rise in oxygen availability 600 million years ago  
coincides with the  Cambrian explosion of life forms~\cite{He_natgeo-2019,Chen_natcom-2015}.  
Oxygen is crucial not only for metabolism, by providing high levels of energy~\cite{SchmidtRohr_acsomega-2020},  
but  also as a morphogen: a  molecule that acts as a signal to organize  cells and organisms   
according to its availability. 
Several lines of research have highlighted the key  role of oxygen
in regulating cell adhesion~\cite{Crossin_cam-2012}, division, differentiation and function~\cite{Fathollahipour_cto-2018}, 
as well as morphogenesis and embryonic development~\cite{Scully_dvp-2016,Simon_nrmcb-2008}.

One prominent mechanism through which oxygen  operates is aerotaxis, 
the ability of motile cells and organisms to bias their movement according to  the surrounding oxygen landscape.  
Aerotaxis is common in bacteria~\cite{Taylor_armb-1999,Morse_bpj-2016,Bouvard_pre-2022} 
and  now emerges as a highly conserved trait among eukaryotes~\cite{Deygas_natcom-2018}, 
that may, for instance,  promote metastatic spread of tumor cells~\cite{Godet_natcom-2019,Semenza_tmm-2012}.
The interplay between oxygen consumption and aerotaxis  
can lead to self-generated gradients  and self-organized patterns,  
as in clumping~\cite{Fenchel_biolrev-2008}, collective migration~\cite{CochetEscartin_elife-2021} or microphase separation of cells~\cite{Carrere_natcom-2023}.  
Aerotaxis is thus one driving force for self-organization of living matter.  

The simplest  instance of aerotactically driven self-assembly  is the aerotactic band.  
The first observations date back to 1881, when Engelmann 
reported  the accumulation of bacteria in the vicinity of the air-water interface~\cite{Taylor_armb-1999}.   
Noting  that bacterial suspensions  in a test tube form a stationary band near the meniscus, 
Beijerinck proposed in 1893 that the underlying mechanism 
is a  preference  for a  specific concentration of oxygen.    
Band formation is not peculiar to the laboratory but expected in many  natural habitats, 
including marine and lake sediments~\cite{Fenchel_biolrev-2008,Brune_femsmr-2000}. 
Indeed, whenever oxygen consumption exceeds the diffusive flux coming from the interface, 
a gradient of oxygen results. 
Microorganisms can then congregate in a horizontal layer, 
whose position reflects the preferred oxygen concentration~\cite{Taylor_tbis-1983}. 
A common behavior  is microaerophily, the preference for oxygen levels much below saturation value. 
Though mostly reported for bacteria,  
aerotaxis and microaerophily are also found in protozoa  such as the ciliate  {\it Euplotes}~\cite{Fenchel_arProtist-1989}. 
Overall, both in terms of habitats and organisms, 
aerotactic band formation  is common in the natural world.   

Aerotactic bands have been investigated experimentally 
in a variety of bacteria~\cite{Baracchini_jpathbact-1959},  
which  includes  microaerophiles {\it Azospirillum brasilense}~\cite{Barak_jbact-1982,Grishanin_jgmb-1991,Zhulin_jbact-1993,Zhulin_jbact-1996} 
and {\it Thiovulum majus}~\cite{Fenchel_microb-1994}, 
obligate aerobe  {\it Bacillus subtilis}~\cite{Baracchini_jpathbact-1959,Wong_jbact-1995}, 
facultative anaerobe {\it Spirochaeta aurantia}~\cite{Greenberg_jbact-1977} and  {\it Shewanella oneidensis}~\cite{Stricker_jrsi-2020},  
aerotolerant anaerobe {\it Desulfovibrio} species~\cite{Krekeler_femsmbec-1998,Eschemann_envmicrob-1999,Fischer_femsme-2006}, 
and {\it  Escherichia coli}~\cite{Ghosh_arxiv-2025}. 
Another class of band forming microorganism is 
that of magnetotactic bacteria~\cite{Frankel_bpj-1997,Smith_bpj-2006,Erglis_mhd-2012,Lefevre_bpj-2014,Popp_natcom-2014}. 
The chain of magnetosomes they synthesize  
acts as a single magnet and orients their body along  Earth's magnetic field~\cite{Lefevre_mmbr-2013,Klumpp_pr-2019,Marmol_rmp-2024}. 
The advantage of such external guidance  to reach the preferred oxygen level 
is still being discussed~\cite{Smith_bpj-2006,Bennet_plosone-2014}.  
Whatever the bacteria species, 
the spatial scales of the band are quite similar: 
the distance to the interface is typically millimetric and 
its width is on the order of  $100\,$\microm.

On the theoretical side, 
modelling of aerotactic bands has provided a basic understanding of the phenomenon~\cite{Mazzag_bjp-2003}. 
The motion of swimming bacteria usually involves persistent  episodes, called runs,  
punctuated by sudden reorientation events~\cite{book_Berg-EColiinmotion,GrognotTaute_comb-2021,Detcheverry_pre-2017}. 
Aerotaxis is achieved
by modulating the reorientation rate as a function of oxygen concentration and gradient.  
The resulting accumulation around a preferred oxygen level   
has been investigated in a number of  approaches, 
none of which, however, is without limitation. 
Individual-based simulations offer a microscopic description~\cite{Codutti_ploscb-2019} 
but the findings may be difficult to generalize beyond the specific system considered. 
Analytical theories based on  Keller-Segel equations~\cite{Keller_jtb-1971,KellerSegel_jtb-1971,Rosen_bmb-1978}
have been extensively developed for chemotactic bands~\cite{Tindall_bmb-2008,Wang_dcdsb-2013}  
but the shallow gradient assumption involved~\cite{Mazzag_bjp-2003,Schnitzer_sympsocgenmb-1990_spe} 
may not hold in aerotactic bands, 
since  the modulation of reorientation rate is not small at the scale of the band.  
Besides, 
the microscopic model of Mazzag \etal~\cite{Mazzag_bjp-2003} has been instrumental 
for the description of the bands~\cite{Smith_bpj-2006,Erglis_mhd-2012,Bennet_plosone-2014,Knosalla_applicamath-2015, Knosalla_jmaa-2019}. 
Nevertheless, motion is restricted to one dimension, 
a convenient assumption justified for magnetotactic bacteria aligned by a strong magnetic field  but an approximation otherwise.  
Finally, the kinetic theory recently proposed~\cite{Ghosh_arxiv-2025} is only solved  numerically.   
Overall, what has been missing is a microscopic but analytical description of  the band  
that while accounting from multidimensional motion,  
is based on the Fokker-Planck equation governing the bacterial distribution.  
Recently, such a framework  has proven a powerful one   
to model rheological properties, including negative viscosity, 
of bacterial suspensions~\cite{Haines_pb-2008,Saintillan_expmech-2010,Vincenti_prf-2018,Saintillan_arfm-2018}.  

In this work, our first objective is to provide a microscopic, Fokker-Planck-based 
description of the aerotactic band. 
Focusing on the steady state, 
we  derive the single equation that connects oxygen concentration  and aerotactic response.   
For two simple aerotactic behaviors, 
we obtain the exact bacterial distributions  
and characterize the band structure. 
Our second objective is to understand analytically the recent observation that 
an aerotactic band of magnetotactic bacteria in a magnetic field 
spontaneously generates  a fluid flow~\cite{Marmol_arxiv-2025}. 
In contrast with the simplified modelling  proposed so far,   
we consider the full interplay 
between aerotactic behavior, oxygen diffusion and hydrodynamic flow.  
We  give  an analytical yet simple prediction for the flow profile 
and show that our description is consistent with the main experimental features. 
Overall, the Fokker-Planck  framework  developed here 
provides a sound basis to model the properties of  aerotactic band, 
and more generally the aerotactic self-assembly of swimming microorganisms.

The remainder of this article is organized as follows. 
Section~\ref{sec:I} focuses on the steady aerotactic band.  
We first write the Fokker-Planck equation that governs the interplay 
between self-propulsion,  rotational diffusion and aerotaxis 
and derive the non-linear third order differential equation that controls the band profile.  
Next, 
we consider  two idealized aerotactic responses that are amenable to an exact solution 
and describe  the symmetric and asymmetric band structure. 
We also discuss how more realistic aerotactic kernels 
influence the band profile and how they might be inferred from experimental data.  
Section~\ref{sec:II} is devoted to the fluid flow observed 
in bands of magnetotactic bacteria~\cite{Marmol_arxiv-2025}. 
We consider in turn the regimes of weak  and strong field. 
Combining  active hydrodynamic stress with an approximate Green function for the fluid flow,  
an elementary expression  is obtained for the flow profile. 
We compare our predictions to experimental findings 
and discuss the connection with microscopic bacterial properties.  
Some perspectives are given in Section~\ref{sec:conclusion}.   
For clarity, the most technical parts of the computation and the large formulas are relegated to Appendices.

\section{The aerotactic band}
\label{sec:I}

In this section, we study a steady aerotactic band.    
We first establish the equation governing the oxygen concentration 
and show that it involves a small dimensionless parameter that results in an inner layer. 
We also present two exact solutions and discuss the connection to experiments.

\subsection{Governing equations}

\subsubsection{Bacteria distribution and oxygen concentration}

\begin{figure*}[t!] 
\center
\includegraphics[width=1\textwidth]{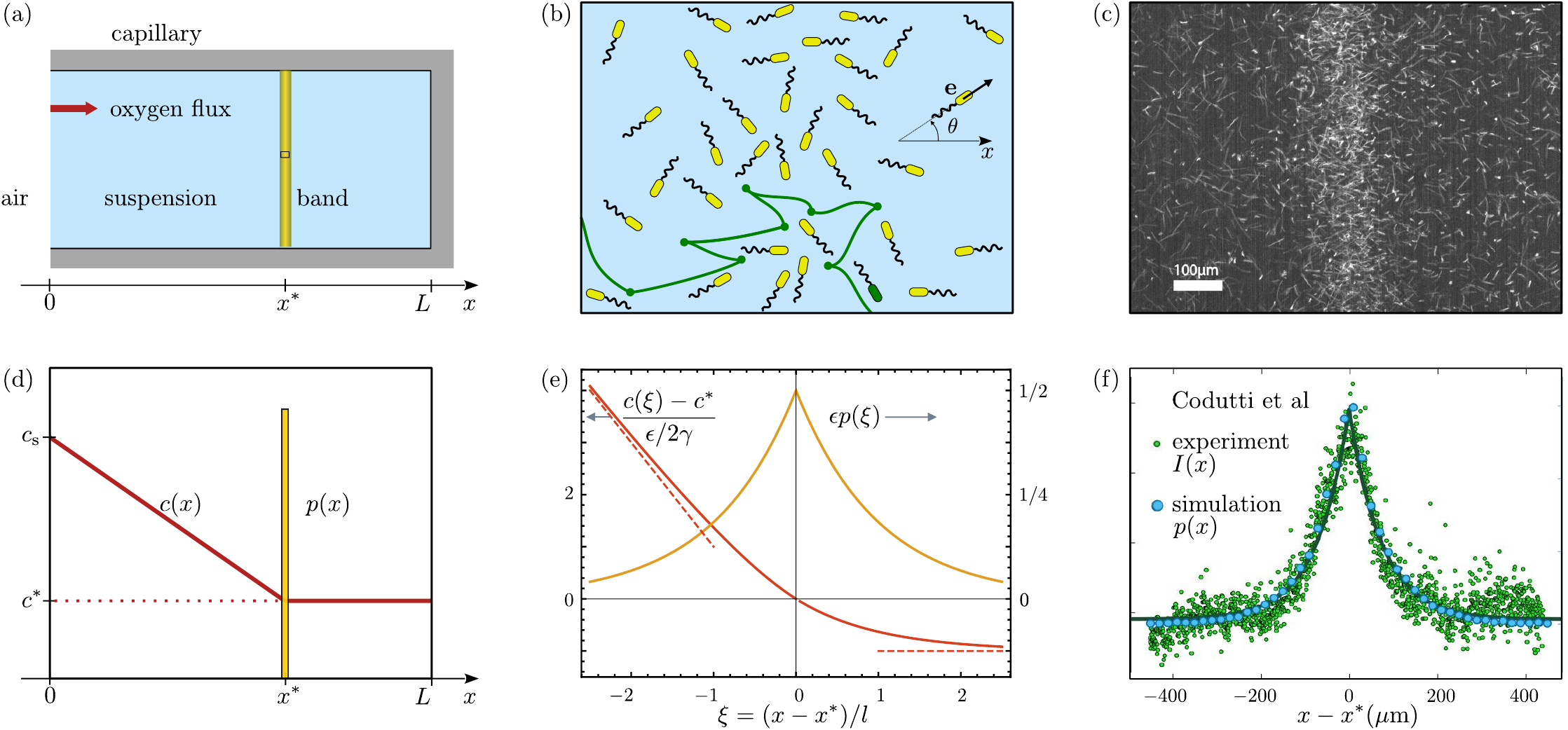} 
\caption{
\capti{The aerotactic band.} 
(a)~Experiment set-up:  
bacteria confined in a capillary form 
a steady band at position~$\xs$ from the air-liquid interface.  
(b)~Microscopic view of the band. Bacteria are not drawn  to scale. 
In the bacterial trajectory (green line), each dot is a tumble. 
(c)~Visualization of the band, as reported in  Ref.~\cite{Codutti_ploscb-2019}, 
 namely  Codutti \etal, {\it PLoS Comput. Biol.} {\bf 15}(12) e1007548 (2009).
(d)~Bacterial density profile and oxygen concentration described at the macroscopic scale. 
(e)~Same quantities described at the microscopic scale. 
The continuous lines correspond to Eq.~\eqref{eq:cin} 
and the dashed lines to Eq.~\eqref{eq:cout} with $\eps=0$.  
(f)~Density profile in the band found by Ref.~\cite{Codutti_ploscb-2019} 
in simulations and in experiments. 
In the latter, 
the grey scale intensity  $I(x)$, in arbitrary unit, is a proxy for local density. 
The line shows a Laplace distribution with a band width $\len=73\,$\microm. 
Panels~(c) and~(f) are adapted from Ref.~\cite{Codutti_ploscb-2019}.  
}  
\label{fig:band}
\end{figure*}

We consider a bacterial suspension confined in a capillary 
with one open end and one closed end~(Fig.~\ref{fig:band}a-c). 
The system may be inhomogeneous along the tube direction ($x$-axis) 
but is assumed invariant in the perpendicular directions, 
thus neglecting the  presence of the meniscus that curves the air-water interface. 
Bacteria swim with constant velocity~$\vo$ 
and a direction that is subject to tumbles and to rotational diffusion with coefficient~$\difrot$.  
Tumble events are instantaneous, governed by a Poisson process and characterized 
by a distribution of turning angle~$h(\thetaturn)$~\footnote{ 
$\thetaturn$ is the angle between direction before tumble and direction after tumble.}. 
For now,      
we assume a given local tumbling rate $\lam(x,\ve)$ 
that depends both on the bacteria position and on orientation, 
as specified by  the unit vector~$\ve$.  
We call~$p(x,\ve,t)$ the probability density to find at time~$t$ 
a bacteria with position~$x$ and orientation~$\ve$. 
For simplicity, 
we consider first a two-dimensional system,  
where bacteria orientation is indicated by the angle~$\theta$~(Fig.~\ref{fig:band}b). 
The Fokker-Planck equation for the probability distribution  $p(x,\theta,t)$ 
can be written as 
\begin{align}
 \partial_t p =& - \vo \costheta \, \partial_x p  + \difrot \partial^2_{\theta\theta} p    \nonumber  \\
 &- \lam(x,\theta)  p + [ \lam(x,\theta)  p] \circ  h,    \label{eq:FP}
\end{align}
Here, the  $\circ$ symbol indicates convolution with respect to orientation, 
that for functions $f(\theta)$ and $g(\theta)$  is defined as
\begin{align}
 f \circ g (\theta)  \eqdef \intpi f(\theta') \, g(\pv(\theta-\theta')) \, \dd \theta', 
\end{align}
with $\pv(\theta) \eqdef \arg e^{i\theta}$ taking value in $[-\pi,\pi[$. 
The first two terms in the right-hand side of Eq.~\eqref{eq:FP} 
account respectively for self-propulsion and rotational diffusion 
while the last  two terms describe tumble reorientation  
with rates that depend on bacterial state.  

We choose a simple form for the tumbling rate. 
In general,  tumbling is a non-Markovian process that involves the whole history of bacteria position $\vcr(t)$. 
For instance, 
linear response theory  postulates~\cite{deGennes_ebpj-2004}
\begin{align}
\lam(t)  = \lamo - \int_{-\infty}^{t} R(t-s) C( \vcr(s) )\, \dd s, 
\end{align}
where $\lamo$ is a base value, 
$R(t)$ a response function and $C$~the concentration field of the species inducing chemotaxis. 
Throughout this work, for analytical tractability,  we assume a tumbling rate of the form   
\begin{align}
 \lam(x,\theta) = \lamo \left( 1 + \krn(x) \costheta \right),  \label{eq:lamxtheta} 
\end{align}
where the dimensionless kernel $\krn(x)$ specifies,  in relative term,  
the spatial modulation of the tumbling rate.    
Equation~\eqref{eq:lamxtheta} involves two approximations. 
First, tumbling is  a Markovian process 
as it involves only the current position and orientation,  
thus discarding history effects  such as response delay.  
Second, 
the directional dependence of the tumbling rate modulation 
reduces to  $\costheta$. 
Such simple cosine dependence  
is found in linear response~\cite{deGennes_ebpj-2004} and 
arises when retaining only the lowest order in a Legendre series expansion~\cite{Schnitzer_pre-1993}.  
More intuitively, 
it reflects the apparent chemical gradient experienced by the moving bacteria: 
for a small displacement $d\vcr$,  the bacteria sees a concentration change 
$d C = d\vcr \cdot \nabla C \sim \ve \cdot \ve_x =  \costheta$, 
since the gradient  is along the $x$-direction.
We note that 
in a recent experimental study of aerotactic bacteria \bcontaminans{}~\cite{Bouvard_pre-2022},   
the local tumbling rate could be measured and  
the two assumptions involved in Eq.~\eqref{eq:lamxtheta} 
were found to be adequate~\footnote{
Reference~\cite{Bouvard_pre-2022} 
assume $\tau = \tauo(1 - k \costheta)$, which is equivalent to our expression, 
for $k$  sufficiently small. 
}. 

Assuming a local tumbling rate  and steady state, 
the Fokker-Planck Eq.~\eqref{eq:FP} admits  the solution~\cite{Schnitzer_pre-1993}: 
\begin{align}
p(x,\theta) = \nor  \exp\left[ - \frac{\Krn(x)}{\kap}  \right],  
\quad 
\kap \eqdef  \frac{\vo \lamoi}{\alb}.  \label{eq:pxtheta}
\end{align}
Here, 
the $\kap$ length is introduced for conciseness, 
$\nor$ is a normalization constant, 
$\al \eqdef \langle \costheta \rangle_{h(\theta)}$ is the mean cosine of turning angle, 
$\alb \eqdef 1-\al$ and 
$\Krn(x)$ is a primitive of $\krn(x)$. 
A noticeable feature of the solution  
is that rotational diffusion does not appear 
and that the distribution of bacterial orientation remains everywhere isotropic. 
As a side comment,  
one can also remark  that Eq.~\eqref{eq:pxtheta} 
takes a form similar to the equilibrium distribution of a Brownian particle 
in a space-dependent force field~$\krn(x)$ deriving from a potential~$\Krn(x)$. 

We now connect the local tumbling rate to the oxygen distribution in the capillary. 
It is postulated that the spatial dependence of $\krn(x)$ 
arises only through the oxygen concentration $c$ and its derivative, 
namely $\krn(x) = \krn(c(x), c'(x))$~\footnote{
The second derivative~$c''(x)$ may be involved as well. 
For higher derivative, the inner layer argument presented in Sec.~\ref{sec:dimless} 
may break down.}.   
The concentration $c(x,t)$ is governed by diffusion and bacterial consumption:
\begin{align}
 \partial_t c = D \partial^2_{xx} c -  \rhom \Lc  p(x,t) \Phicons(c),   \label{eq:oxdif}
\end{align} 
with $D$ the diffusion coefficient of oxygen, 
$\rhom$ the mean number density of bacteria in the suspension, 
$\Lc$ the length of th capillary~\footnote{Said otherwise,  $L$ is the tube length. 
It is not the capillary length associated to the air-liquid surface tension 
and that would be involved in the description of the meniscus.} 
and $p(x,t)$ the probability density of bacteria at position $x$. 
$\Phicons(c)$ gives the individual consumption of a bacteria at concentration~$c$.  
Assuming steady state and combining Eq.~\eqref{eq:pxtheta} and Eq.~\eqref{eq:oxdif} 
lead to a single equation on oxygen concentration
\begin{align}
\frac{c'''}{c''} - \frac{c' \Phicons'(c)}{\Phicons(c)} = - \frac{\krnetc}{\kap}.  
\end{align}
In the following, we consider only the simplest choice for~$\Phicons$:  
the individual consumption is a  constant~$\qo$, 
which yields the aerotactic band equation 
\begin{align}
 - \kappa \,c''' = c'' \, \krnetc.    \label{eq:eqcdim}
\end{align}
This third-order differential equation is 
subject to three boundary conditions. 
First, 
at the air-liquid interface near the capillary open end, 
the liquid is saturated in oxygen:  $c(0)=\csat$ with $\csat$ the saturation value. 
Second,  the oxygen flux entering the tube is entirely consumed by cells, 
which gives $c'(0)= - \qo \rhom L/D$. 
Third, the oxygen flux vanishes at the closed end of the capillary, $c'(L)=0$. 

Equation~\eqref{eq:eqcdim} is the main result of this section. 
Given an aerotactic behavior specified by the kernel $\krnetc$,  
it determines the oxygen concentration profile that exists 
within the capillary in steady state.  
The bacterial density distribution follows from Eq.~\eqref{eq:oxdif} 
as $p(x) = c''(x) D/\qo\rhom L$. 
Though derived for a two-dimensional system, 
Eq.~\eqref{eq:eqcdim} also applies in other dimension. 
In three dimension, the convolution term in Eq.~\eqref{eq:FP} 
is replaced by an azimuthally symmetric convolution on the unit sphere~\footnote{See Appendix A in Ref.~\cite{Detcheverry_pre-2017} for details.}
but the outcome is identical.  
In one dimension, 
Eq.~\eqref{eq:FP} still holds if taking  $\alb=1$, as detailed in Appendix~\ref{sec:app:1D}. 
The one-dimensional model assumes that motion occurs only along the capillary axis 
and that tumbles are perfect reversal. 
Such a situation can be experimentally realized if 
magnetotactic bacteria  swimming with a symmetric run-reverse  strategy~\footnote{We assume forward and backward modes have strictly identical properties, including velocity and reversal rate.}
are aligned along the capillary axis by a strong magnetic field.

\subsubsection{Dimensionless form and interior layer}   
\label{sec:dimless}

It is useful to express the problem in dimensionless form. 
To do so,  the capillary length~$L$ is taken as unit length  
and the oxygen saturation value~$\csat$ as unit concentration, giving the governing equations 
\begin{subequations}
\label{eq:eqdimless} 
\begin{align}
- \eps \, c'''(x) & = c''(x) \krnetc,                                              \label{eq:eqdimlessa}  \\
c(0)            & = 1,                     \quad c'(0)=-\gami,    \quad c'(1)=0,   \label{eq:eqdimlessb}  \\
p(x)            & = \gam c''(x),                                                   \label{eq:eqdimlessc}
\end{align}
\end{subequations}
where from now on, all quantities are dimensionless, 
unless  mentioned otherwise. 
The problem involves the two dimensionless parameters
\begin{align}
\eps \eqdef \frac{ \vo \lamoi}{\alb L},     \qquad    
\gam \eqdef \frac{\csat D}{ \qo \rhom L^2}.            \label{eq:epsgam}
\end{align}
In the simplest case that will be studied below (Sec.~\ref{sec:toysensing}),  
$\eps$ and $\gam$ controls respectively the width and position of the band. 
Let us estimate their order of magnitude. 
As regards oxygen property, 
the diffusivity  is  $D=2\,10^{-9}\,$m$^{2}\,$\si 
and  $\csat = 0.2\,$mol$\,$m$^{-3}$ at saturation~\cite{booksec_xyl-2014}. 
Experimental systems~\cite{Mazzag_bjp-2003,Elmas_bmcm-2019,Codutti_ploscb-2019} 
involve 
a bacterial velocity~$\vo = 10-50\,$\microm$\,$\si, 
a tumbling rate  around $\lamo = 1\,$s$^{-1}$  
and a capillary length~$\Lc=4-40\,$mm.  
Finally, 
we consider a typical bacteria density $\rhom = 10^{14}\,$m$^{-3}$~\cite{Mazzag_bjp-2003,Bouvard_pre-2022}  
and an oxygen consumption $\qo = 10^{-18}\,$mol$\,$\si~\cite{Mazzag_bjp-2003}. 
With those values, 
$\gam$ pertains to the range $2\,10^{-3} - 0.3$ and can be close to unity, 
whereas 
$\eps$ falls in the range $2\,10^{-4} - 10^{-2}$ and is always much below unity.

The smallness of the dimensionless parameter $\eps$ 
has a generic consequence: 
the aerotactic band is, mathematically speaking, an interior layer~\cite{book_Holmes-IntroPertMeth}.  
Because $\eps$ appears as a factor of the highest derivative in the differential Eq.~\eqref{eq:eqdimlessa}, 
one can expect a small region where the concentration exhibits very strong variations. 
Such an interior layer is similar to the boundary layers 
commonly found in fluid mechanics~\cite{book_White-ViscousFluidFlow,book_Pozrikidis-IntroTheoCompFluidDyn} 
but it occurs in the interior of the domain. 
The band width is of order $\eps \Lc$,  much narrower than the capillary. 
In the limit $\eps \to 0$, 
the bacteria distribution and oxygen concentration profiles  
are very simple~(Fig.~\ref{fig:band}d):  
bacteria accumulate at a given position, say $\xs$, 
with a probability distribution $p(x)$ approaching $\delta(x-\xs)$.  
The oxygen concentration is piecewise linear  since $c'''(x)=0$, 
and Eq.~\eqref{eq:eqdimlessb} leads to 
$c(x) = 1- x/\gam$ for $x<\xs$ and 
$c(x) = c(\xs) \eqdef \cs$ for $x>\xs$~\footnote{
An experimental oxygen profile consistent with those expectations is given in Fig.~2  of Ref.~\cite{Zhulin_jbact-1996}. 
}. 
For a finite value of~$\eps$, 
such a description is an approximation valid at the macroscopic scale and outside the band.
To resolve the microscopic structure of the band,  
one needs to specify the tumbling kernel $\krnetc$.

\subsection{Exact solutions for band structure}

\subsubsection{Toy sensing: the Laplace band}
\label{sec:toysensing}

A commonly observed aerotactic behavior is the drive toward 
a preferred oxygen concentration $\cs$~\cite{Fenchel_microb-1994,Zhulin_jbact-1996,Fischer_femsme-2006,Lefevre_bpj-2014,Ghosh_arxiv-2025}. 
In particular,  microaerophile bacteria seek oxygen levels  well below the saturation ($\cs \ll \csat$).  
Probably the simplest way to encapsulate the preference toward $\cs$  in the tumble rate kernel is to write 
\begin{align}
\krnetc= \ko \sign \left(\cs-c\right),    \label{eq:toysensing}
\end{align}
with $0<\ko<1$ a dimensionless constant. 
Whatever the oxygen level or gradient, the bias in  tumbling rate has the same magnitude 
but its sign indicates whether the local concentration is above or below the  target value. 
For convenience, whenever the concentration dependence of the kernel 
reduces to $\sign \left(\cs-c\right)$, we speak of binary sensitivity. 
As discussed in Appendix~\ref{sec:app:toysensing}, 
the toy model embodied in Eq.~\eqref{eq:toysensing} 
is only an approximation %
but it has the chief advantage of tractability. 

With most choices of kernel $\krnetc$, 
the third-order differential equation governing the oxygen concentration (Eq.~\eqref{eq:eqdimlessa})  
is nonlinear and challenging to solve.  
With binary sensitivity, however, it admits an exact solution. 
We applied a singular perturbation analysis~\cite{book_Verhulst-MethApplSingPert,book_Holmes-IntroPertMeth} 
which exploits the smallness of $\eps$ dimensionless number~\footnote{
The solution for arbitrary $\eps$ can also be written but differs by exponentially small terms such as $\exp(-\xs/\eps)$, 
which are negligible in practice.}. 
The concentration outside the band is given by an outer expansion, which at lowest order in $\eps$, reads as
\begin{subequations}
\label{eq:cout}
\begin{align}
\cout(x) =& 1   - \frac{x}{\gam},    \;\;\;\quad \mbox{for} \:  x<\xs \eqdef \gam (1 - \cs) +\frac{\eps}{2},     \label{eq:couta} \\
\cout(x) =& \cs - \frac{\eps}{2\gam},     \quad \mbox{for} \:  x>\xs.                                            \label{eq:coutb}
\end{align}
\end{subequations}
On the other hand, the concentration within the band is given by the inner expansion 
\begin{subequations}
\label{eq:cin}
\begin{align}
\cin(x)   &=        \cs + \eps \, \Psi\left( \frac{x-\xs}{\eps} \right),                   \\
\Psi(u) &\equiv   \frac{1}{2 \gam} \left( - 2 u H(-u) + e^{- |u|} -1  \right),   
\end{align}
\end{subequations}
with $H$ the Heaviside function. 
The bacterial density is 
\begin{align}
p(x) = \gam \cin''(x) = \frac{1}{2\eps} \exp\left[ -\frac{|x-\xs|}{\eps} \right],     \label{eq:rhoin}
\end{align}
which is valid everywhere. 
As expected, a Dirac function is recovered in the limit $\eps \to 0$.  
Figure~\ref{fig:band}e illustrates the oxygen concentration 
and the bacterial density profile at the scale of the band. 

To discuss the band structure, 
we express  the bacterial probability distribution
in dimensional quantities~\footnote{Dimensional results are recovered with the substitution $x \to x/\Lc$, $p \to \Lc p$ and $c \to c/\csat$.} 
\begin{align}
 p(\xs+x,\theta) =  \frac{\piso(\theta)}{2 \len}\exp\left[ -\frac{|x|}{\len} \right],  
 \quad 
 \len \eqdef \frac{\vo \lamoi}{\alb \ko},    \label{eq:pLaplace}
\end{align}
where $\piso(\theta)=1/2\pi$ (resp. $(\sin \theta)/2$) denotes an isotropic distribution of orientation in dimension $d=2$ (resp. $d=3$). 
By analogy with the Laplace distribution for a continuous random variable, 
the aerotactic band given by Eq.~\eqref{eq:pLaplace}  can be  called the Laplace band. 
The exponential decay may be understood intuitively 
when remembering that $\krnetc$ may be seen as a force (see Eq.~\eqref{eq:pxtheta}).  
In the present case, the force is a constant on each side of the band ; 
the profile is thus exponential  
as in Perrin's experiment with colloids sedimenting under gravity~\cite{Perrin_ancp-1909}.  
As can be expected, 
the band width~$\len$ increases with mean run length $\leno \eqdef \vo \lamoi$   
and decreases when modulation in tumbling rate is stronger or 
when tumbling events are efficient in reversing the direction of motion. 

Could the toy sensing model embodied in Eq.~\eqref{eq:toysensing} 
provide an adequate description  of some bacterial bands?    
Shown in Fig.~\ref{fig:band}f  is an intensity profile 
measured by Codutti \etal~\cite{Codutti_ploscb-2019} 
for an aerotactic band of magnetotactic bacteria {\it Magnetospirillum gryphiswaldense}. 
Using the deviation of intensity from base value as a proxy for local bacterial density   
yields the density profile shown with green dots.    
The data is perfectly compatible with a Laplace profile and gives a band width 
$\len = 73\,$\microm. 
In  our model 
this would imply $\tauo \ko \alb = 3\,$s 
and would provide an estimate for $\ko$, 
if the mean run time $\tauo$ and $\alb$ could be measured directly from individual trajectories~\footnote{In this analysis, 
we have neglected the presence of an imposed magnetic field along the capillary axis  
because at least to the naked eye, 
the bacterial trajectories suggest  no strong alignment along the magnetic field.}.    
Besides, Ref.~\cite{Codutti_ploscb-2019} shows that 
simulations of an agent-based model also yield a Laplace profile~\footnote{
Given the parameters $\vo=14\,$\microm$\,$\si, $\alb=2$, $\ko=1$, 
our model would predict a band width $l \simeq 14\,$\microm, 
well below the value of $75\,$\microm{} reported for simulation. 
It is unclear which differences between the model and the simulation explains this discrepancy. 
}, 
as also shown in Fig.~\ref{fig:band}f. 
Somewhat unexpectedly, the simplest model of aerotactic behavior,  
and one of the few solvable cases, 
may serve as a useful approximation for at least one type of real bacteria band~\footnote{
As a first approximation, the aerotactic band reported for {\it Thiovulum majus} 
in Ref.~\cite{Fenchel_microb-1994} may also be described as a Laplace band.}.

\begin{figure}[t!] 
\centering
\includegraphics[width=0.42\textwidth]{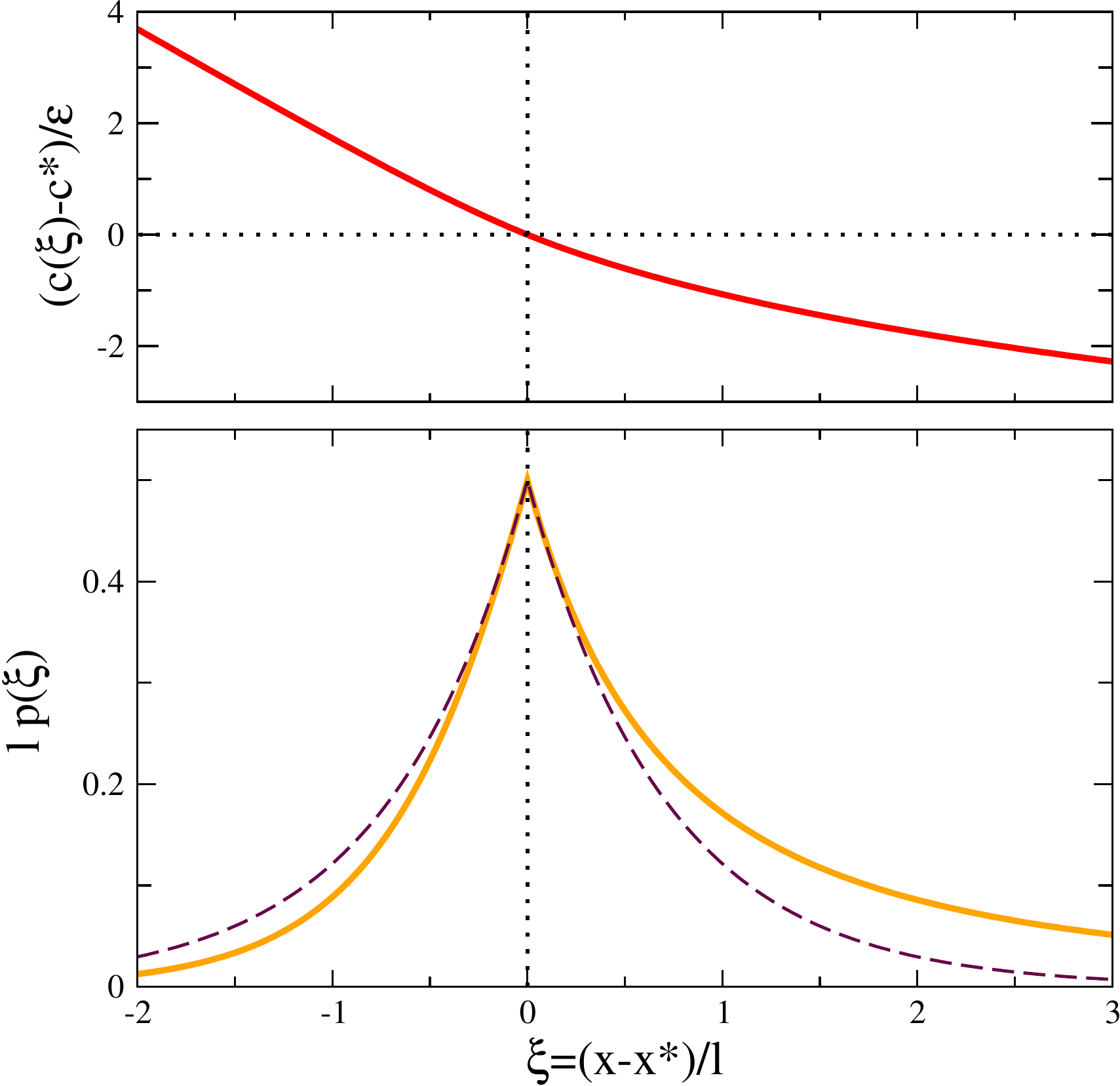}  
\caption{
\capti{Aerotactic band with linear sensing.}
Oxygen concentration profile (top) and bacterial density (bottom), 
as given by Eq.~\eqref{eq:linearbandc} and Eq.~\eqref{eq:linearbandp} respectively.  
The dashed line is  an exponential approximation valid near the band center, 
shown for comparison.}  
\label{fig:lin}
\end{figure}

\subsubsection{Linear sensing: an asymmetric band}
\label{sec:linearsensing}
We now consider linear sensing, i.e. an aerotactic response 
proportional to the perceived oxygen gradient, as is the case in linear response. 
Keeping the assumption of binary sensitivity, the rate modulation is now 
\begin{align}
 \krnetc  =  -\frac{c'(x)}{\cop} \, \sign\left(\cs-c\right),      
\end{align}
where  $\cop>0$ is  a characteristic concentration gradient introduced to make $\krn$ dimensionless. 
Proceeding as above, one arrives at 
\begin{align}
\eps \,c''' = c'' c' \sign(c-\cs), \quad \eps \eqdef \frac{\cop  \leno}{\alb \csat},    
\end{align}
with $\leno \eqdef \vo \lamo^{-1}$ the mean run length in the absence of modulation.  
Note that the definition of $\eps$ depends on the context (see Eq.~\eqref{eq:epsgam}) 
but the same notation is kept  because this parameter 
plays a similar role in the governing equation. 

The dimensionless parameter $\eps$ is again a small one. 
To reach this conclusion, consider $\Delta c \approx \cop \leno$, 
the typical concentration change seen during one run. 
Assuming $\Delta c \approx \csat$ would imply that 
a rate modulation of order unity is reached only for extremely strong gradients, 
in which the maximal change possible 
occurs over the mean run length.  
Since a significant aerotactic response is typically observed 
for gradients that are not so extreme,  
one can expect $\Delta c$ significantly below $\csat$, 
leading to $\eps \approx \Delta c/\csat \ll 1$. 

The band with linear sensing is  amenable to an exact solution.  
The concentration profile solving the nonlinear differential equation can be written as
\begin{subequations}
\label{eq:linearbandc}
\begin{align}
 c(x)        &= \cs + 2 \eps\, \Psi(\xi),                          \quad \xi = \frac{x-\xs}{l}, \;\; l \eqdef 2\gam \eps, \\
 \Psim(\xi)  &= \ln \left[ \cosh(\xi - \xione)/\sqrt{2} \right],   \\
 \Psip(\xi)  &= -  \ln \left[ 1 + \xi/\sqrt{2} \right],
\end{align}
\end{subequations}
with the numerical constant $\xione \eqdef \ln(1+\sqrt{2})$. 
Here and below, the $\pm$ sign  
indicates the negative and positive side  of the band. 
Besides, the band is centered at position 
\begin{align}
 \xs  = \gam (1 - \csat) + 2 \gam \eps \ln \left[ 2/(1+1/\sqrt{2}) \right]. 
\end{align}
Finally, the bacteria probability density is 
\begin{subequations}
\label{eq:linearbandp}
\begin{align}
p_{-}(\xi) &= \frac{1}{l}  \sech^2 (\xi - \xione),   \\
p_{+}(\xi) &= \frac{1}{l}  (\sqrt{2}+\xi)^{-2},   
\end{align}
\end{subequations}
In writing the equations above,  
the band is assumed to be sufficiently thin and located far away from the capillary ends,  
which is always the case  for $\eps$ small enough. 

With respect to the Laplace band, 
the band formed with linear sensing has a number of new features. 
On the negative side, 
the decay is exponential both at small and large distance but with distinct length scales~\footnote{
$p_{-}(\xi) \simeq   \exp(- \sqrt{2} |\xi| )/2 $ for small $\xi$ while 
$p_{+}(\xi) \simeq 4 \exp(- 2 |\xi - \xione| ) $ for large $\xi$.}. 
On the positive side, the decay of density $ p_{+} \sim \xi^{-2}$ 
is algebric and thus much slower. 
Indeed, with the concentration constrained between $0$ and $\cs$, which is often small itself, 
the gradient $c'(x)$ is rather weak there. 
Finally, the band width in real units is $l = \eps \gam L = \leno \cop /|c'(0)|$,     
where $c'(0)$,   the concentration gradient at the capillary open end, 
gives a typical gradient value in the system.   
If the unmodulated mean run length $\leno$ is known, 
an experimental observation of the band width 
would give access to the microscopic parameter~$\cop$.   

\subsection{Discussion} 

\subsubsection{Continuous aerotactic kernel}
\label{sec:continuouskernel}

\begin{figure}[t!] 
\center
\includegraphics[width=0.42\textwidth]{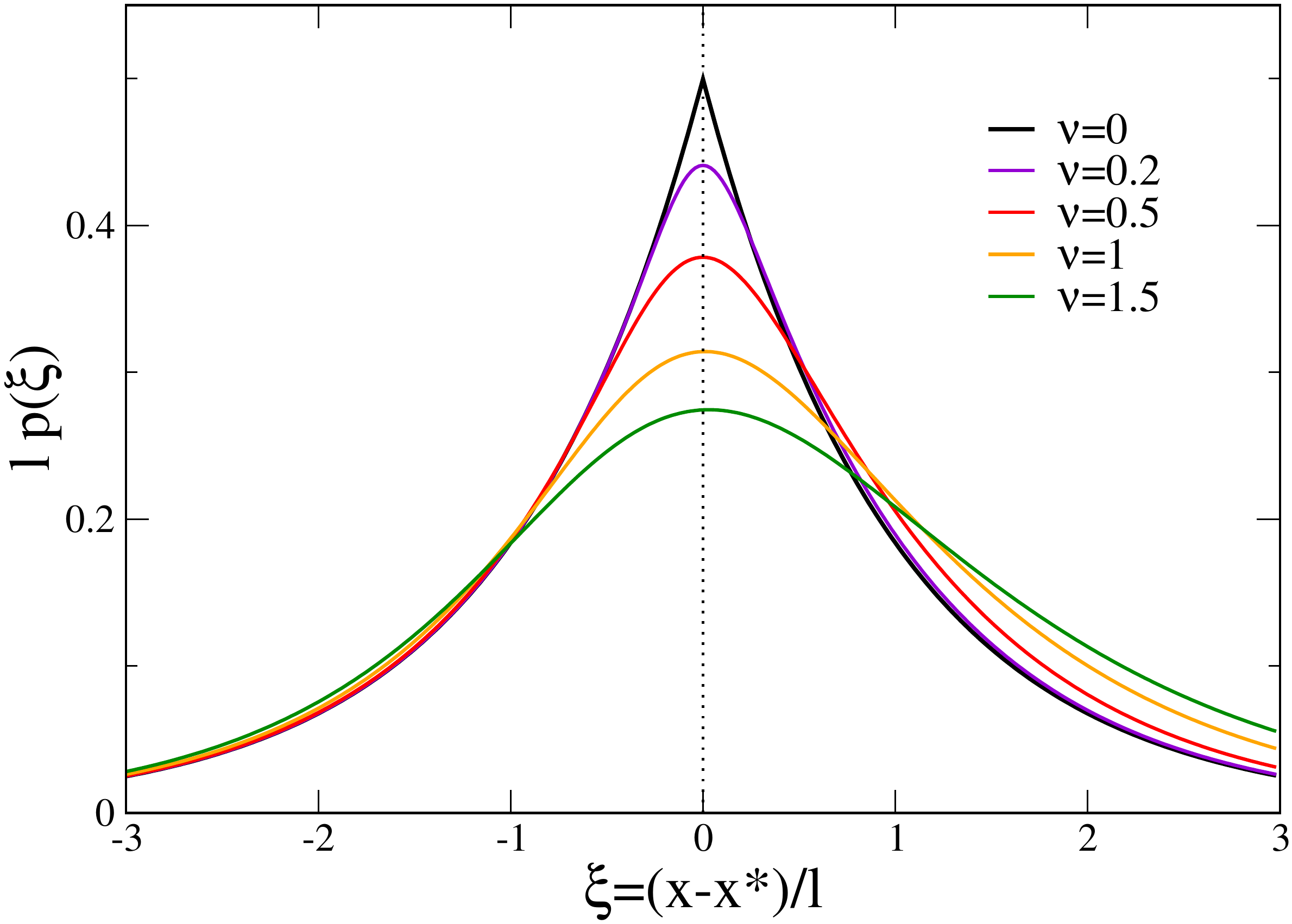}  
\caption{
{\bf Smooth band profile with a continuous kernel.}
Bacterial density computed by numerical resolution of Eq.~\eqref{eq:eqdimless} 
with increasingly gradual aerotactic kernels,   
as specified by~$\nupar$ values in Eq.~\eqref{eq:kernelsmooth}. 
The black curve ($\nupar=0$) corresponds to the binary sensitivity of Eq.~\eqref{eq:toysensing}. 
}  
\label{fig:smooth}
\end{figure}

Binary sensitivity has the advantage of tractability 
but is likely, for most bacteria, to remain an idealization.  
One  particular feature that may appear unphysical  
is the discontinuity in aerotactic response.  
Instead, one could expect a smooth crossover between the two saturation values, 
for at least two reasons. 
At the single bacterial level, 
the chemical network underlying aerotaxis may lead to a sigmoid response. 
At the population level,  
cell variability could induce, rather than a fixed value,  a distribution of preferred concentration~$\cs$, 
which would result in an effective aerotactic kernel that is continuous~\footnote{
The effective kernel would be  $\krnetc$ averaged over the probability distribution $P(\cs)$ of $\cs$.}. 
To illustrate the effect of a smooth aerotactic kernel, let us consider the simplest choice 
\begin{align}
 \krnetc = \ko  \tanh \left[ \frac{\cs-c}{\nupar \cs}  \right],     \label{eq:kernelsmooth}
\end{align}
where any sigmoid function could be used in place of the hyperbolic tangent 
and the dimensionless parameter~$\nupar$ specifies the relative deviation with respect to $\cs$ 
beyond which the response  saturates.  
The resulting band profiles, computed numerically, are illustrated in Fig.~\ref{fig:smooth}. 
In comparison to the Laplace band, two modifications are apparent in the profile:  
rounding and asymmetry. 
Specifically, the cusp point at the band center is replaced by a smooth profile 
rounded on a characteristic length  fixed by~$\nupar$.   
As regards asymmetry, 
the departure from the Laplace profile is more pronounced on the low-oxygen side of the band 
because with $c$ restricted to positive value,
the difference $\cs-c$ remains  so small there that the saturation value is not reached. 
We note finally that for $\nupar$ as large as $1/2$, and outside the central part of the band, 
the density profile  remains close to that obtained for $\nu=0$. 
In that sense, binary sensitivity may remain a convenient approximation. 

\subsubsection{Inference of aerotactic kernel}

Band profiles that are neither exponential nor symmetric 
have been observed for several bacteria species in experiments~\cite{Lefevre_bpj-2014}. 
Can we infer from the data the underlying aerotactic behavior?  
If the kernel depends only on oxygen concentration~$c$, 
the answer is positive. 
Indeed, exploiting Eq.~\eqref{eq:eqcdim} and Eq.~\eqref{eq:eqdimlessc}  gives in this case
\begin{align}
 \krn (c) =  - \kappa \, \frac{p'(x(c))}{p(x(c))},      \label{eq:inference}
\end{align}
where $x(c)$ is the position at which concentration $c$ is reached~\footnote{
A bijective relation between $x$ and $c$ should exist since $c(x)$ at steady state should be strictly decreasing, 
at least in the presence of bacteria.}. 
If the bacteria density profile and oxygen concentration 
can be measured simultaneously and with sufficient accuracy, 
Eq.~\eqref{eq:inference} allows to estimate the aerotactic kernel. 
A similar conclusion holds 
when the dependence in gradient is linear and 
$\krn$ of the form $c' \hat{k}(c)$,  with $\hat{k}(c)$ the function to determine. 
When, on the other hand, the aerotactic response involves 
a more complex relation with  $c$ and $c'$, 
the spatial variation is not sufficient to disentangle each dependence. 
Experiments with imposed gradients~\cite{Kalinin_bpj-2009,Adler_loac-2012} might then be necessary.

\subsubsection{Relationship with Keller-Segel approach}

As a final point, 
we discuss the relation between our microscopic description  
and a macroscopic modelling based on Keller-Segel type equation~\cite{Keller_jtb-1971,KellerSegel_jtb-1971}. 
Such a framework has been extensively used to describe 
chemotactic band, aggregation and patterns~\cite{Tindall_bmb-2008,Wang_dcdsb-2013}. 
In this approach, the bacterial flux $J(x)$  is written as 
\begin{align}
 \frac{J(x)}{\rhom L} =  - \difb  p'(x) + \va p(x),   \label{eq:J}
\end{align}
where $\difb$ is the long-time diffusion coefficient (diffusivity) of bacteria 
and $\va$ is the macroscopic aerotactic drift velocity.  
How  is the drift $\va$ related to the aerotactic kernel~$\krnetc$ 
of the microscopic description? 
The relationship is a simple one.  
In the absence of tumble rate modulation,   
the bacteria diffusivity is given by $d \difb = \vo^2/(\difrot' + \lamo \alb)$~\cite{Lovely_jtb-1975,Taktikos_plosone-2013}, 
where $d$ is the space dimension and $\difrot' \eqdef (d-1)\difrot$. 
For a steady band, Eq.~\eqref{eq:J} gives $\va = \difb p'(x)/p(x)$ and 
\begin{align}
 \va = - \frac{\vo \krnetc}{d(1+\difrot'\lamoi/\alb)}. 
\end{align}
In the particular case where rotational noise is negligible~\footnote{
This is often the case. 
With typical values $\difrot = 0.1\,$\si{} and $\lamoi=1\,$s, 
the product $\difrot \lamoi$ is small compared to unity.}, 
one recovers $\va = - \vo \krnetc/d $ as derived in Ref.~\cite{Bouvard_pre-2022}. 
Thus, the macroscopic drift velocity appearing in a macroscopic Keller-Segel description 
is simply proportional to the tumbling kernel. 
In that sense, the two descriptions may be considered as equivalent. 

That being said, one should note that  the correspondence  may not hold beyond the steady state. 
More importantly, the equivalence may also break down when considering magnetotactic bacteria.  
How to extend the Keller-Segel equation in this case is far from straightforward 
because the presence of a magnetic field induces drift and anisotropic diffusivity. 
Therefore, one must resort to a microscopic modelling, 
that accounts for the distribution of bacteria position and orientation. 
The natural framework   to do so is  the  Fokker-Planck equation,  
to which we now come back, this time accounting for magnetic effects.

\section{Field-induced fluid flow in a band of magnetotactic bacteria}
\label{sec:II}

\begin{figure}[t!] 
\center
\includegraphics[width=0.47\textwidth]{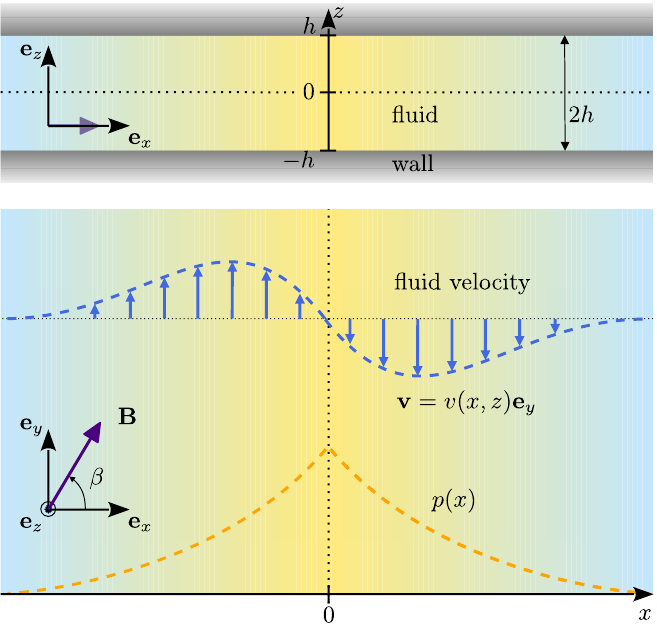}  
\caption{
\capti{Field-induced fluid flow in an aerotactic band of magnetotactic bacteria (MTB)}.   
Schematic of the experimental findings of Ref.~\cite{Marmol_arxiv-2025}.  
The suspension is confined in a capillary, 
with walls at vertical position $z=\pm h$, $h$ being the half-thickness. 
In the presence of a magnetic field $\vB$ whose orientation lies in the $x-y$ plane,  
a spontaneous  flow  is generated along the band direction,  
but with opposite signs on each side. 
The bacterial density~$p(x)$ and the fluid velocity profile~$v(x)$  
are shown pictorially. 
The capillary used in experiments has lateral dimensions much larger than the band width 
and  our model  assumes infinite extent.    
}  
\label{fig:bandB}
\end{figure}

We consider an aerotactic band of magnetotactic bacteria (MTB), 
submitted to a uniform and constant magnetic field $\vB$, as depicted in Fig.~\ref{fig:bandB}. 
Recent experiments~\cite{Marmol_arxiv-2025} report the spontaneous flow of the suspension,  
together with a mesoscopic and approximate model that points to the role of hydrodynamic active stress. 
We develop here an exact, microscopic, Fokker-Planck description,  
with the aim of understanding both the band structure and the field-induced fluid flow. 

\begin{figure*}[t!] 
\center
\includegraphics[height=0.27\textwidth]{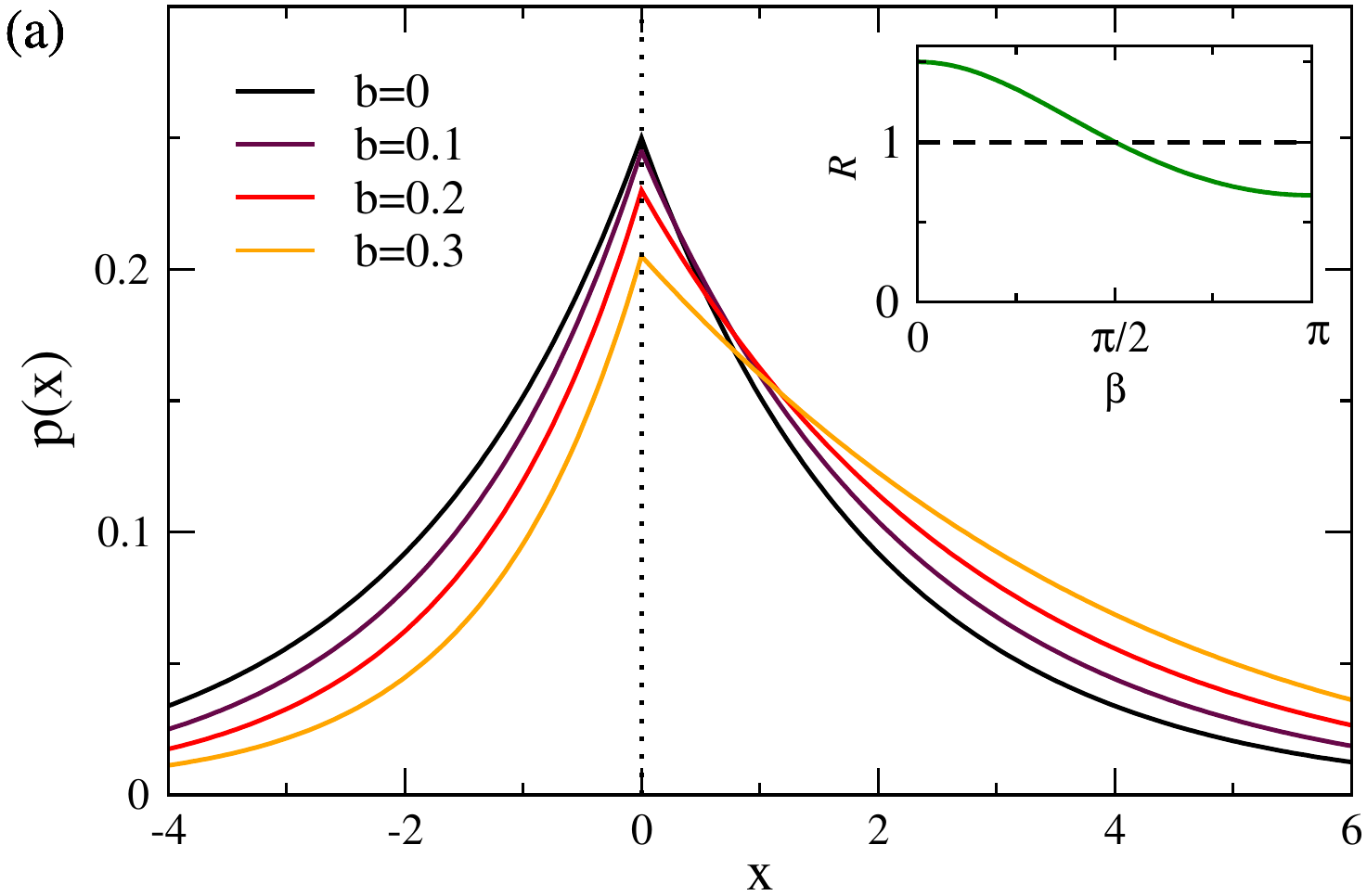}  \hspace*{1.2cm} 
\includegraphics[height=0.27\textwidth]{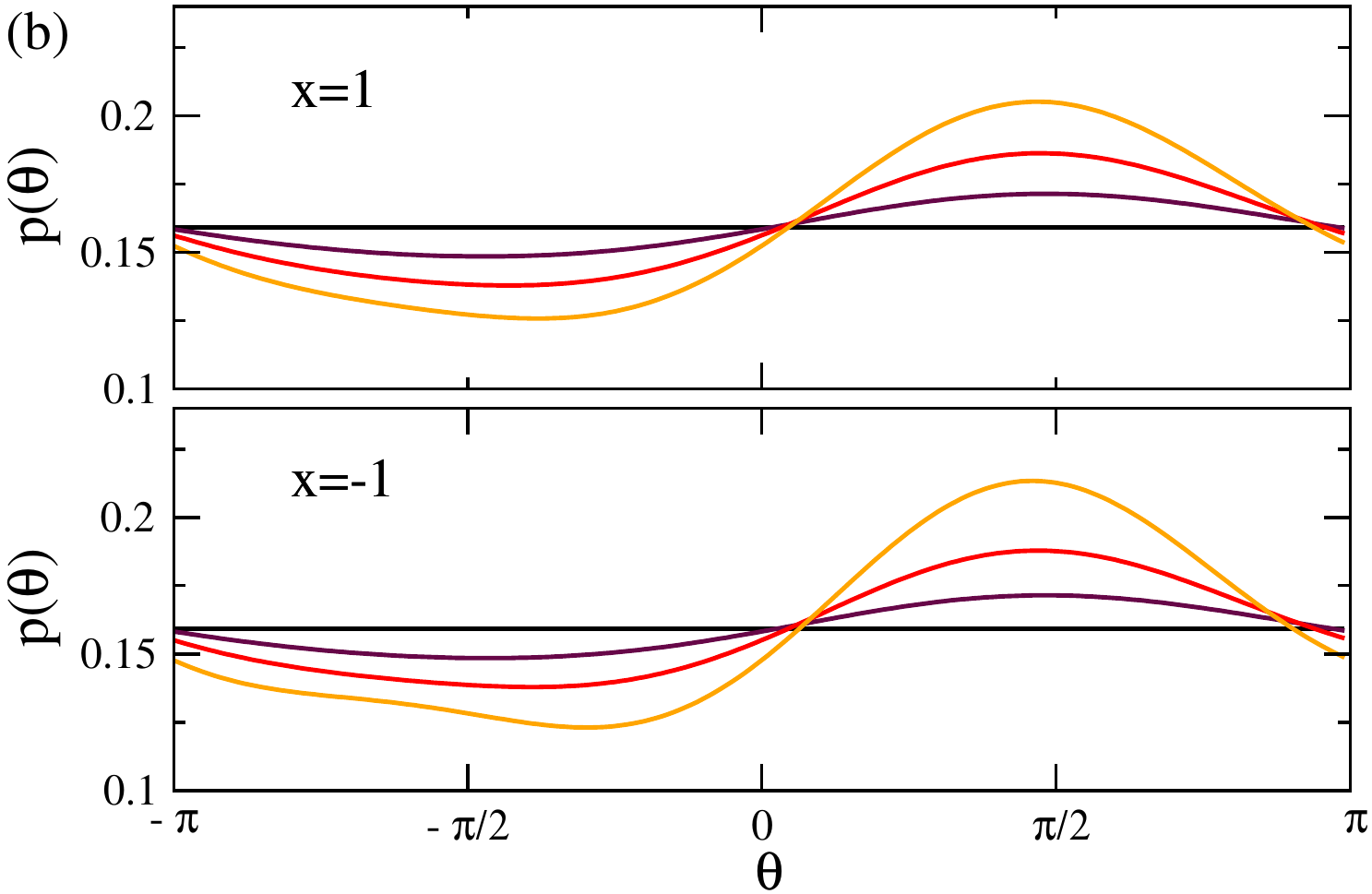} 

\caption{
{\bf Structure of the aerotactic band of magnetotactic bacteria under weak magnetic field.} 
(a)~Bacteria density profile~$p(x)$. 
The ratio $R$ defined by Eq.~\eqref{eq:R} is shown in the inset for $b=0.1$.  
(b)~Orientation distribution $p(\theta)$, computed at position $x=1$ and $x=-1$.  
Curves are plotted from Eqs~\eqref{eq:pxthetafieldA}-\eqref{eq:pxthetafieldC} 
valid at second order in~$b$. 
Parameters: the tumbling modulation is $k=1/2$ and the field orientation given by $\bet=\pi/4$.  
}  
\label{fig:lowfield}
\end{figure*}

Throughout this section,  
we will assume that bacteria have  a two-dimensional orientation  within the field plane, 
a restriction made for tractability's sake.  
Working now exclusively at the microscopic scale of the band, 
we take $\lamoi$ as unit time and $\vo \lamoi$ as unit length, 
unless mentioned otherwise.  
The magnetic field is parallel to the $x-y$ plane  
and its orientation with respect to the $x$-axis  is specified with the $\bet$~angle~(Fig.~\ref{fig:bandB}). 
The Fokker-Planck Eq.~\eqref{eq:FP} is then supplemented  on the right-hand side 
with an additional term $b \, \partial_\theta (  \sin(\theta-\bet) p)$, 
where the dimensionless magnetic field $b$ is  defined as
\begin{align}
 b \eqdef \frac{B}{\Bc} \eqdef \frac{\mo B}{\Gamo \lamo}. 
\end{align}
The characteristic field $\Bc$ is constructed from the magnetic moment $\mo$ and rotational friction coefficient $\Gamo$  of the MTB. 

In the following, 
we will consider only the toy sensing model, as defined by Eq.~\eqref{eq:toysensing}. 
Such a  choice  is made for tractability, 
since, in the absence of field, the Laplace band has the simplest structure. 
Besides, we will mostly investigate the situation 
where the magnetic field is a small perturbation to the band. 

\subsection{Bacteria distribution at weak field}   

We start with the particular case of a field perpendicular to the band ($\bet=0$), 
where the probability density of bacteria keeps  a simple form:
\begin{subequations}
\begin{align}
2 \pi p(x,\theta) &= \frac{H(x) e^{-x/\lp} +  H(-x) e^{x/\lm}}{\lp+\lm},   \label{eq:pxthetabeta0}  \\
l_{\pm}           &\eqdef \frac{1}{\alb \kom \mp b},   
\end{align}
\end{subequations}
whose validity requires a sufficiently low field $b \leqslant  \alb \kom$, 
The band becomes asymmetric, with a different decay length on each side.   
Yet, the distribution of orientation is still isotropic, there is no tendency to align along the field. 
Such a screening effect of the field by aerotaxis is surprising at first sight 
but can be understood as follows. 
When the orientation of bacteria is isotropically distributed, 
the field term  $b\, \partial_\theta (  \sin(\theta-\bet) p)$ 
reduces to $b \costheta \,p$, 
which has exactly the same form as the tumbling rate modulation.
Therefore the effective force associated to binary sensitivity aerotaxis 
is simply shifted by the magnetic field, 
yielding an asymmetric but still exponential, Perrin-like decay. 

For an arbitrary orientation of the field, 
the Fokker-Planck equation becomes much more challenging to solve. 
Accordingly, we assume from now on that tumbles are isotropic~\footnote{ 
The distribution of turning angle is then $h(\thetaturn)=1/2\pi$ and $\al=1-\alb=0$.}.  
Furthermore, we consider the limit of weak field, $b \ll 1$,  
and seek a second-order expansion in $b$ as
\begin{align}
 p(x,\theta) =  p_0(x,\theta) + b p_1(x,\theta) + b^2 p_2(x,\theta), 
\end{align}
where $p_0$ is the solution in the absence of field. 
Exact expressions for $p_1$ and  $p_2$ can be found, as detailed in Appendix~\ref{sec:app:lowfield}.
The solution method relies on Laplace transform for position and Fourier series for orientation 
and exploits symmetry and normalizability constraints on $p(x,\theta)$. 
The end result for the bacteria distribution can be written as 
\begin{subequations}
\begin{align}
 \frac{p(x,\theta)}{p(x)} =&  \frac{1}{2\pi} \biggr( 1 + b \sin \bet \sin \theta           
                                             - \frac{b^2}{4} \sin(\bet)  \times  \nonumber   \label{eq:pxthetafieldA}  \\
                           & \quad \Bigr[   3 \sin(\bet - 2 \theta) + \sin(\bet + 2 \theta) \Bigr] \biggr),     \\  
\frac{p(x)}{p_0(x)}  =&  1 + b x \cos \bet + \frac{b^2}{2} (x^2-\frac{2}{\kom^2} )  \cos^2 \bet,  \label{eq:pxthetafieldB}     \\
p_0(x)               =& \frac{\kom}{2} e^{-\kom x},                                               \label{eq:pxthetafieldC}   
\end{align}
\end{subequations}
where all expressions hold at second order in~$b$. 
Here and in the following,  
the band center is now taken as the origin and is located at $x=0$, 
as in Fig.~\ref{fig:bandB}. 
Equations~\eqref{eq:pxthetafieldA}-\eqref{eq:pxthetafieldC} are valid for the positive side of the band ($x>0$) ;   
the distribution on the negative side can be obtained by the substitutions: 
$x\to-x$, $\theta \to \theta+\pi$, $\bet \to \bet+\pi$. 
For simplicity, 
rotational diffusion has been neglected ($\difrot=0$) 
but the solution at finite $\difrot$ is also available, as given in Appendix~\ref{sec:app:lowfield}.  

We now discuss how a weak magnetic field modifies the aerotactic band.   
From symmetry, it is sufficient to consider a field orientation $\bet$ in the interval $[0,\pi/2]$. 
For $\bet=0$, 
the density profile is different than in the no-field case but the orientation distribution is unaffected. 
For $\bet=\pi/2$, the situation is reversed.  
At intermediate value of $\bet$, 
both density profiles and orientation are modified by the field,  as illustrated in Fig.~\ref{fig:lowfield}. 
The band is asymmetric, with  decay lengthes given by $l_\pm = 1/(\kom \mp b \cos \bet)$. 
Denoting as $p_{\pm}$ the amount of bacteria on each side, one gets the ratio 
\begin{align}
 R \eqdef \frac{p_+}{p_-} = \frac{\kom+ b \cos \bet}{\kom -  b \cos \bet},    \label{eq:R}
\end{align}
which is shown in the inset of Fig.~\ref{fig:lowfield}a. 
As regards the orientation, 
one could expect a growing peak around $\theta=\bet$ that would indicate alignment with the field. 
However, at this level of approximation, 
the dominant term is the first order correction of the form $\sin \theta$, 
and except for its magnitude, 
the modulation of orientation (Fig.~\ref{fig:lowfield}b) appears  only weakly dependent on field direction. 

In addition to modifying band structure, 
an other effect of the field is to induce a drift motion of bacteria. 
For convenience, let us introduce, for any quantity $F(x,\theta)$, the average with respect to orientation: 
\begin{align}
\avg{ F(x,\theta)} &\eqdef \frac{ \intpi p(x,\theta)  F(x,\theta) \dd \theta }{p(x)}.  
\end{align}
The intrinsic  velocity of bacteria, that is relative to the surrounding fluid,  
has the average components:  
\begin{subequations}
 \begin{align}
\avg{\vint_x} &\eqdef   \avg{ \costheta } = 0,                         \\
\avg{\vint_y} &\eqdef   \avg{ \sin \theta } = \frac{b}{2} \sin \beta,  
\end{align}
\end{subequations}
with the velocity expressed in $\vo$ unit. 
The bacteria thus drift parallel to the band everywhere, 
which has no further consequence since we assume invariance in the $y$-direction.   

\begin{figure*}[t!] 
\center
\includegraphics[height=0.27\textwidth]{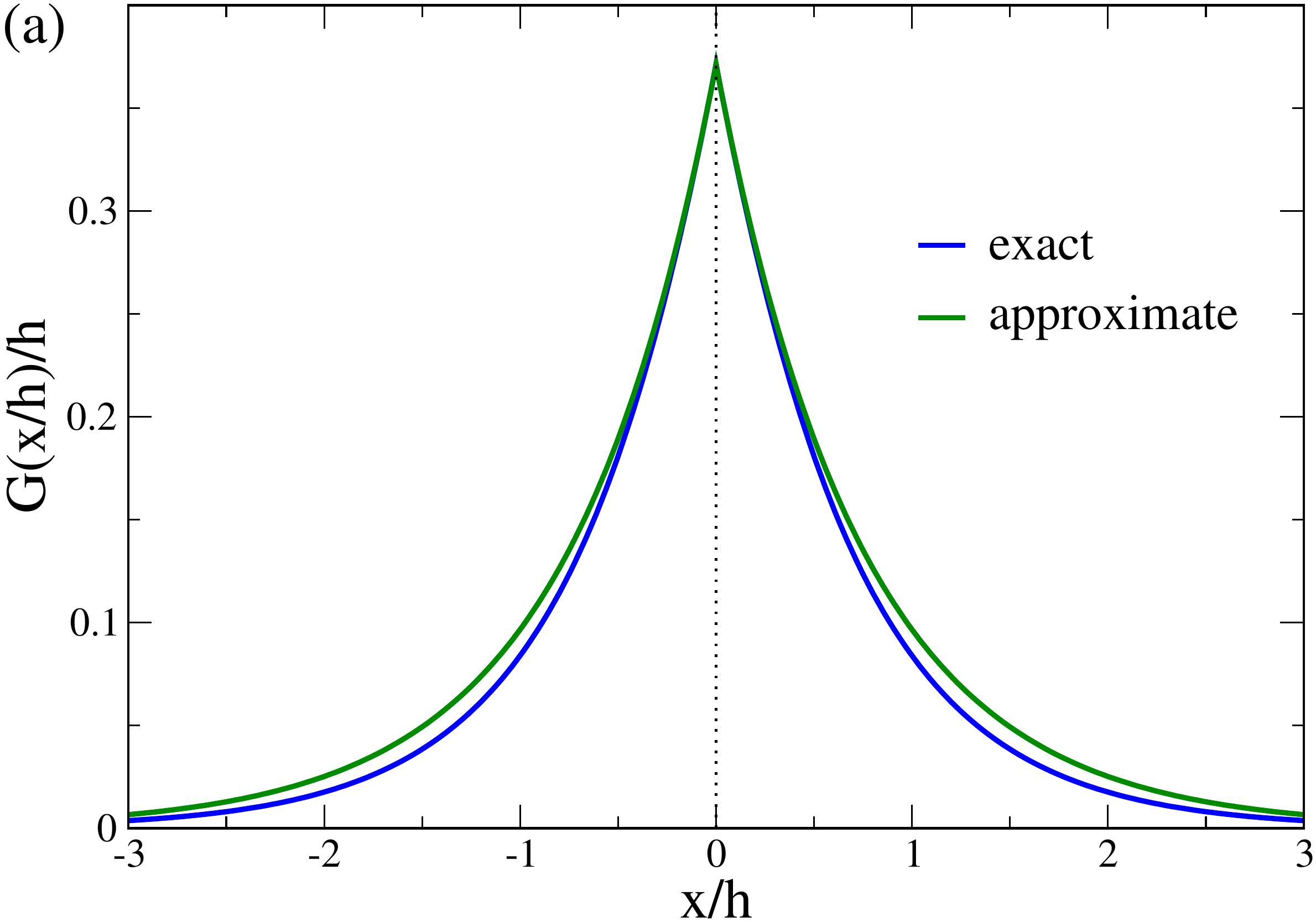}  \hspace*{1.2cm} 
\includegraphics[height=0.27\textwidth]{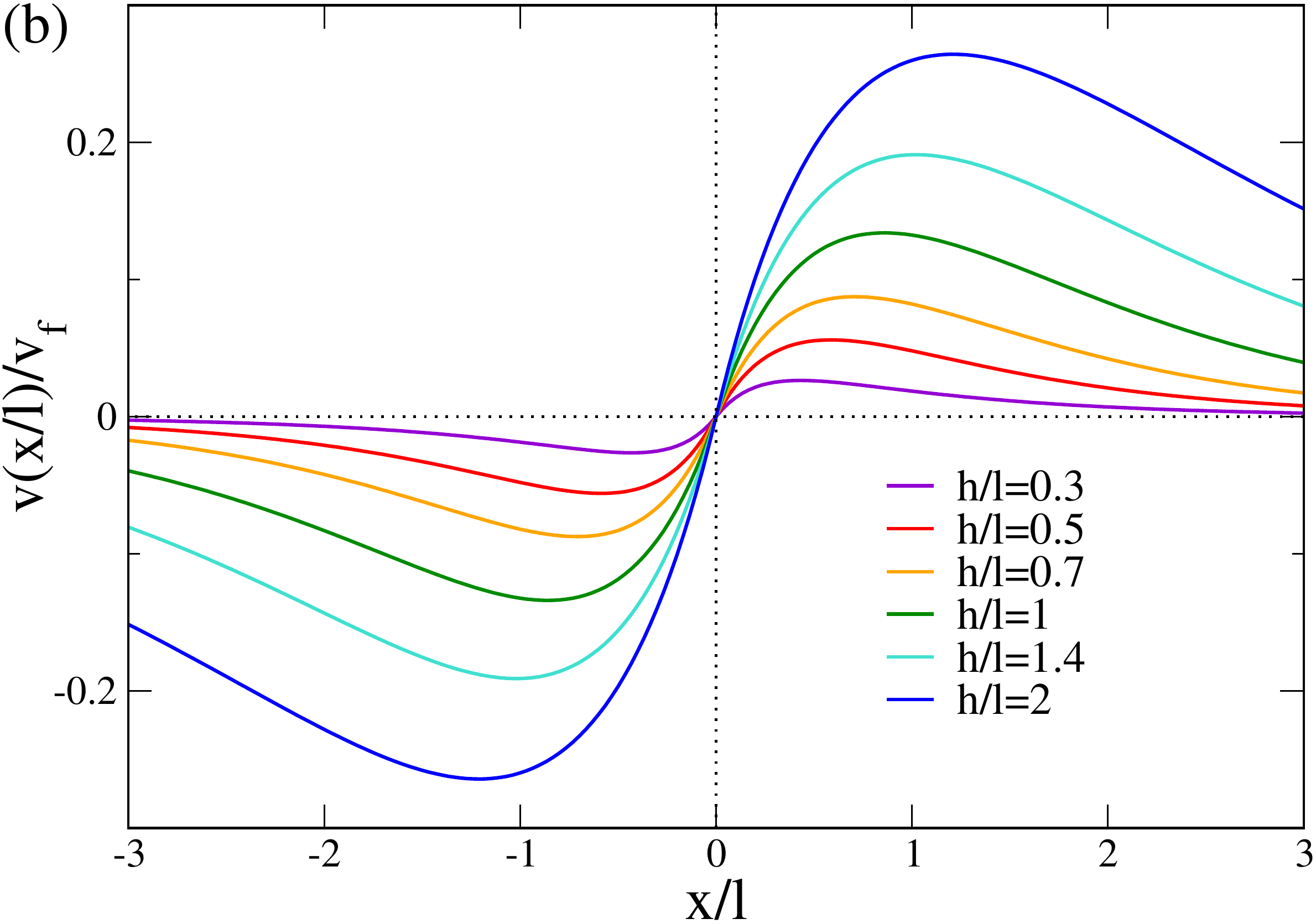} 
\caption{
{\bf Field-induced flow in an aerotactic band of magnetotatic bacteria.} 
(a)~Green function for the hydrodynamic problem. 
The exact expression $G(x)$ and its approximation $\Gapp(x)$ 
are given by Eq.~\eqref{eq:Gexact}   and Eq.~\eqref{eq:Gapp} respectively. 
(b)~Velocity profile induced by a Laplace band of width~$l$ 
in a capillary of half-thickness~$h$. 
The velocity  is normalized by $v_\mathrm{f} \eqdef \Vscale   \sin(2\bet) b^2/8$. 
}  
\label{fig:flow}
\end{figure*}

\subsection{Fluid flow}

\paragraph*{Active stress and flow problem.} 
Let us now examine the effect of bacteria self-organisation 
on the surrounding fluid. 
We focus here on the active contribution to stress induced by bacterial swimming, 
other contributions will be discussed below in Sec.~\ref{sec:discB}.  
The active stress tensor is~\cite{Saintillan_expmech-2010,Saintillan_arfm-2018}   
\begin{align}
\tensTa  \eqdef \sigo  \rho  \Lc  \, p(x) \left[ \avg{\ve \ot \ve}  - \frac{\tensI}{2}  \right].   \label{eq:Ta}
\end{align}
Here, 
$\sigo$~is the force dipole moment of the bacteria, assumed fixed, 
$\ve \eqdef [\costheta, \sin \theta]$~the orientation vector, 
$\tensI$~the identity matrix and  direct product is implied.   
Exploiting Eqs.~\eqref{eq:pxthetafieldA}-\eqref{eq:pxthetafieldC}, one finds 
\begin{align}
  \tensTa     & = \sigo  \rho  \Lc \frac{b^2}{16} \, p_0(x) 
  \begin{bmatrix}
  -4 \sin^2 \beta &      \sin(2\beta) \\
   \sin(2\beta)   &  4 \sin^2 \beta
  \end{bmatrix},                                                         
\end{align}
The resulting force (volumic) density is then 
\begin{align}
 \vfa \eqdef \nabla \cdot \tensTa 
 = 
 \sigo  \rho  \Lc  \frac{b^2}{16}  p_0'(x)
 \begin{bmatrix}
  -4 \sin^2 \beta  \\
   \sin(2\beta)    
  \end{bmatrix}.                  \label{eq:vfa}
\end{align}
The fluid velocity $\vv(x,y,z)$ is governed by flow incompressibility and 
by the Stokes equation, respectively
\begin{align}
  \nabla \cdot \vv =0,  \qquad 
 \eta \Delta \vv + \vfa = \nabla \pres, 
\end{align}
with $\eta$ the suspension dynamic viscosity, $\Delta$ the Laplacian and $\pres$ the pressure. 
As regards boundary conditions, 
the suspension is confined between top and bottom walls at height  $z= \pm h$, 
where a no-slip condition applies. 
Besides, the flow also vanishes far from the band, for $x \to \pm \infty$.  

\paragraph*{Green function and approximation.} 
For simplicity,  
the fluid flow is approximated as unidirectional and parallel to the band, 
namely $\vv = v(x,z) \,\vey$, 
and as everywhere else, invariance is assumed along the band ($y$-axis). 
The Green function $G(x,z)$ associated to the flow problem is the solution of~\footnote{The minus sign is included for convenience 
so as to make the Green function positive.}
\begin{align}
 \Delta G(x,z) = - \delta(x), \label{eq:eqG}
\end{align}
and describes the flow induced  by a punctual forcing
applied at the origin~\footnote{
We assume everywhere that the bacteria distribution is invariant in the $z$-direction 
and thus neglect surface effects induced by the capillary walls.}. 
To solve Eq.~\eqref{eq:eqG}, it is convenient to work with the Fourier transform with respect to the $x$ position 
and variable~$q$. 
For a function $f(x)$, its Fourier transform $f(q)$ satisfies 
\begin{subequations}
\begin{align}
f(q)   &\eqdef             \int_{-\infty}^\infty    e^{ i q x} f(x)  \dd x,  \\
f(x)   &= \frac{1}{2\pi}   \int_{-\infty}^\infty    e^{-i q x} f(q)  \dd q, 
\end{align}
\end{subequations}
where, to keep light notations,  
the nature of the function is indicated by the name of the variable. 
Using $\Delta = -q^2 + \partial^2_{zz}$  in $(q,z)$ variables 
and the no-slip condition at the wall, the Green function $G(q,z)$ is found to be
\begin{align}
 G(q,z) &= \frac{1}{q^2} \left[  1 - \frac{\cosh(q z)}{\cosh(qh)} \right].    \label{eq:Gqz)}  
\end{align}
In the following, 
we focus on the velocity in the mid-plane because it is representative of the local flow 
and thus consider $G(x) \eqdef G(x,z=0)$. 
 
The Green function $G(x)$ can be obtained explicitly in terms of hyperbolic and polylog functions, 
as reported in Appendix~\ref{sec:app:G}. 
However, to make analytical progress easier, we will use instead the  approximation 
\begin{align}
 \Gapp(x) = \frac{\lamhy}{2} \exp \left[ - \frac{|x|}{\lamhy} \right], \quad \lamhy \eqdef \frac{8 \catalan }{\pi^2} h,   \label{eq:Gapp}
\end{align}
where $\catalan \simeq 0.915$  is Catalan's constant. 
This approximation, illustrated in Fig.~\ref{fig:flow}a, 
is exact at first order in a small-$x$ expansion and is thus accurate near the origin. 
Elsewhere, 
the difference $|\Gapp(x) - G(x)|$ does not exceed $4\%$ of the maximal value $G(0)$. 
In view of its simplicity, 
Eq.~\eqref{eq:Gapp} is remarkably faithful 
and at a minimal loss in accuracy, will allow to express  the flow profile in extremely simple terms.

\paragraph*{Flow profile.}
For an arbitrary forcing $\fa(x)$, 
the induced fluid flow is obtained through the convolution
\begin{align}
 v(x) = - \int_{-\infty}^\infty  G(x-x')\,\fa(x')\, \dd x'.   \label{eq:convol}  
\end{align}
The velocity profile is smoothed over a length $\lamhy \sim h$ 
which is to be compared to the typical scale $\la$ over which $\fa$ varies. 
In the limiting case where $h \ll \la$, 
the Green function approaches $\lamhy^2 \delta(x)$.  
The response of the fluid is then local, 
with a velocity  proportional to the  local forcing,  
as in a Hele-Shaw cell or in Darcy law for flow in porous media~\cite{book_ghp-PhysHydro}.  
Thus, in a capillary much thinner than the band ($h \ll l$), 
the flow profile would simply be proportional to the derivative of the density profile~$p_0'(x)$, 
as indicated by Eq.~\eqref{eq:vfa}. 
The limit of thin capillary, however, does not apply in experiments.  
Instead, the capillary thickness is typically larger than the band width 
and is the characteristic length scale over which the velocity profile varies. 
The fluid response is then non-local. 

We finally make a prediction for the  fluid flow. 
The unperturbed density profile is that of a Laplace band, 
$p_0(x) = 1/(2l) \exp(-|x|/l)$ with the width $l \eqdef \vo \lamoi/k$ in real units. 
Combining  Eq.~\eqref{eq:convol} with the approximate Green function Eq.~\eqref{eq:Gapp} 
yields  the velocity profile 
\begin{align}
 v(x)  &= \Vscale   \sin(2\bet) \frac{b^2}{8} \frac{\sign(x)}{l^2/\lamhy^2-1}  \left[  e^{-|x|/l} - e^{-|x|/\lamhy} \right],  \label{eq:vx}
\end{align}
where $l \neq \lamhy$~\footnote{
In the particular case $l = \lamhy$, the velocity profile is $v(x)  = \Vscale  \, \sin(2\bet) x e^{-|x|/l} /2l$.} 
and we introduced the velocity scale 
\begin{align}
\Vscale \eqdef \frac{\pi}{2} \frac{ \sigo \rho L}{\eta}.  \label{eq:Vscale}
\end{align}
Equation~\eqref{eq:vx} is the main result of this section. 
As illustrated in Fig.~\ref{fig:flow}b, 
the thicker the capillary, the faster the fluid flow and the broader the velocity profile. 
To be more quantitative, 
one can focus on the maximal velocity reached at position $\xm$~\footnote{
The position of the maximum is given by $\xm/l = r \log r/(r-1)$, with $r \eqdef \lamhy/l$. 
In the thin  capillary limit ($h \ll l$), $\xm = \lamhy |\log r|$. 
In the thick capillary limit ($h \gg l$), $\xm = l \log r$.
Up to a logarithmic factor, 
$\xm$ is governed by the smallest length among  $\lamhy$ and $l$.}
and with value  
\begin{subequations}
\begin{align}
\vmo        &=  \Vscale  \sin(2\bet) \frac{b^2}{8} \rondF(\lamhy/l), \qquad \mbox{for\ } b \ll 1,  \\
\rondF(u)  &\eqdef  \frac{u^{(2-u)/(1-u)}}{1+u}. 
\end{align}
\end{subequations}
The function $\rondF(u)$ has simple limiting behaviors: 
$\rondF(u) = u^2$ for  $u \ll 1$ and $\rondF(u)=1$ for $u \gg 1$~\footnote{
Note also the particular value  $\rondF(1)=(2e)^{-1}$. }. 
Accordingly, in the thin capillary regime  ($h \ll l$), 
the maximal velocity  would follow $\vmo \sim \lamhy^2 \sim h^2$, 
a dependence similar to that found in a simple Poiseuille flow~\cite{book_ghp-PhysHydro}. 
In the opposite limit of thick capillary ($h \gg l$), 
the maximal velocity $\vmo$ reaches a plateau set by the velocity scale~$\Vscale$.

\subsection{Discussion}
\label{sec:discB} 

\subsubsection{Simplifying assumptions}  

The model developed so far relies on a number of simplifications. 
In particular, 
it predicts a fluid flow but starting from a Fokker-Planck equation where the effects of flow are absent. 
Besides, only the active contribution to stress tensor was accounted for. 
Here we detail all terms that were not considered so far 
and discuss why they can be discarded or included  within the present framework. 

In the presence of fluid flow,   
the Fokker-Planck equation involves two additional fluxes 
that account for advection and rotation of the bacteria. 
Advection by the fluid yields a flux $\partial_y(v(x) p(x,\theta) )$, 
which vanishes by invariance in the $y$-direction. 
Approximating the bacteria as a rigid particle,  
rotation induced by the flow  can be described by 
Jeffery's equation~\cite{Jeffery_prsa-1922,Bretherton_jfm-1962,Junk_jmfm-2007}, 
\begin{align}
 \dot{\ve} = (\tensI - \ve \ot \ve) \cdot (\tensW + \consB \tensE) \cdot \ve.   \label{eq:jeffery}
\end{align}
Here, 
$\dot{\ve}$ is  the time derivative of $\ve$, 
$\tensE \eqdef (\nabla \vv +\nabla \vv^T)/2$ and  $\tensW \eqdef (\nabla \vv -\nabla \vv^T)/2$, 
where $T$~denotes the transpose, 
are respectively the rate-of-strain and vorticity tensors, 
and $\consB$~is the Bretherton constant. 
For a spheroidal particle of aspect ratio~$r$,  
$\consB = (r^2-1)/(r^2+1)$, with in particular  
$\consB =1$ for an infinitely thin rod and  $\consB=0$ for a sphere.  
Applying Eq.~\eqref{eq:jeffery} to the present system  
leads an  orientation governed by the equation $\dot{\theta} = \left[ 1 + \consB \cos(2\theta) \right] v'(x)/2$, 
as for Jeffery's orbits in simple shear flow~\cite{Petrie_jnnfm-1999,Ishimoto_jpsj-2023}. 
The corresponding flux in the Fokker-Planck equation is thus 
$\nabla_\ve \cdot ( \dot{\ve} \, p(x,\ve) ) = \partial_\theta ( \dot{\theta} \, p(x,\theta) )$. 
As everywhere so far, 
we consider the weak field limit, with terms of at most second order in dimensionless field $b$. 
Because the fluid velocity appears already at second order, 
$p(x,\theta)$ enters only through its zero-order approximation $p_0(x,\theta)$, 
whose orientation is isotropic. 
The Jeffery flux thus reduces to 
$\partial_\theta ( \dot{\theta} \, p_0(x,\theta)  ) = -  \consB v'(x) \sin(2\theta) p_0(x,\theta)$.  
Such a term, however, 
makes the Fokker-Planck equation nonlinear because $v(x)$ is itself proportional to $p(x,\theta)$~\footnote{
The relation between fluid velocity and bacterial density  is 
complicated  since the former results from a convolution between  the latter's derivative and the hydrodynamic Green function.}.  
To avoid the intricacy of nonlinearity in an already complex problem, we fix  $\consB=0$, 
Such an approximation is legitimate if we choose  $\consB \simeq 0$, 
and thus restrict the model to quasi-spherical swimmers. 

Let us now examine the contributions to the stress tensor 
that supplement the active term considered so far.  
The Brownian stress $\tensTBr = 2 \rho L p(x) \kB T ( \avg{ \ve \ot \ve} - \tensI/2)$, 
with $\kB$~the Boltzmann constant and $T$~the temperature, 
has a form similar to the active stress of Eq.~\eqref{eq:Ta} 
and can thus be included as a correction to the force dipole moment~$\sigo$. 
The viscous stress~$\tensTv$ induced by the presence of particles, assumed rigid,  
is given by~\cite{Hinch_jfm-1973,Hinch_jfm-1976,book_Larson-StrRheoComplexFluids}
\begin{align}
\frac{\tensTv}{2\eta \volfrac} = \pA  \avg{\ve \ot \ve \ot \ve \ot \ve} \!:\! \tensE  +  
                               \pB  \left[  \avg{\ve \ot \ve} \cdot \tensI + \tensI \cdot  \avg{\ve \ot \ve} \right]  +  \pC  \tensE, 
\end{align}
with $\volfrac$ the bacteria volume fraction.  
The parameters~$\pA$, $\pB$ and  $\pC$  depend only on particle shape 
and are given in Ref.~\cite{Hinch_jfm-1972} 
in the spheroidal case.   
For a quasi-spherical shape with aspect ratio $r= 1 + \varepsilon$, 
a small-$\varepsilon$ expansion leads to $\pA \sim \varepsilon^2$, $\pB \sim \varepsilon$ and $\pC =5/2$ at lowest order. 
The last term is thus not negligible but 
because $\volfrac$ and $\tensE$ are both linear in bacterial density in a low density expansion, 
it appears only at second order in density 
and can thus be neglected in the dilute limit.

Finally, the last contribution to the stress tensor 
comes from the magnetic field acting on the MTB dipoles. 
The magnetic stress tensor reads as~\cite{Ilg_jcp-2002,Cunha_pf-2024} 
\begin{align}
\tensTm = \frac{\rho}{2} L p(x) \mo B \, \avg{ \veB \ot \ve -  \ve \ot \veB}, 
\end{align}
where $\veB \eqdef \vB/B$ is the unit vector giving the field direction.  
A computation at second order in $b$ gives 
$\vfm \eqdef \nabla \cdot \tensTm = \fm \vey$, 
where the magnetic force density in dimensionless units is 
$\fm = - \mo \Bc \rho L (b^2/8) p_0'(x) \sin(2\bet)$. 
This contribution turns out to be proportional to the active contribution given by Eq.~\eqref{eq:vfa}. 
The effect of magnetic stress can thus be included  at low field 
by simply replacing $\sigo$ with $\sigo - 2 \mo \Bc$. 

To summarize, 
our description involves the following series of assumptions: 
dilute limit, two-dimensional orientation of bacteria, no rotational diffusion, isotropic tumbles,
aerotactic toy sensing with binary  sensitivity, 
low magnetic field,  unidirectional flow and  quasi-spherical shape. 
Within this framework, our description is exact. 

\subsubsection{Strong field limit} 

After investigating the weak field regime, 
we turn briefly to the limit of strong magnetic field. 
Here we do not carry out a perturbative analysis on the bacterial distribution. 
To obtain an approximate estimate, 
we simply consider that all bacteria are perfectly aligned with the field~\footnote{
The probability distribution is thus proportional to $\delta(\theta - \beta)$.}  
and evaluate the resulting fluid flow. 
To be consistent with our assumptions so far, 
it is still assumed that while the force dipole is constant, 
the direction of motion can still be reversed~\footnote{ 
This is possible for instance with biflagellated bacteria. 
In contrast,  a monoflagellated bacteria performing run-reverse by changing motor direction  
would alternate between puller and pusher mode, which would require a different framework.}. 
As the motion occurs exclusively in the field direction,  
we can apply the one-dimensional model of Appendix~\ref{sec:app:1D}. 
The only change required is that the effective velocity orthogonal to the band is now $\vo \cos \bet$. 
The density profile is thus exponential with decay length~$\luni \cos \bet$,  
where~$\luni \eqdef \vo \lamoi/\kom$. 

Within this approximation, 
the band width~$\luni \cos \beta$ vanishes for~$\bet \to \pi/2$. 
Such a behavior arises because the model postulates 
strictly identical bacteria, infinite sensitivity to local concentration and instantaneous response, 
absence of fluctuations in the band position, etc, all conditions that are likely not verified in experiments. 
Besides, as seen above, 
the typical scale over which the flow profile varies is the largest length among~$\lamhy \simeq h$ and~$\len$. 
The former is usually larger than the latter in experiments, 
implying that the vanishing of band width has little effect on  the flow profile. 
For all these reasons, we do not consider further the band narrowing phenomenon. 

It is then straightforward to retrace the steps above to derive the active stress tensor and velocity profile. 
Equation~\eqref{eq:vx} still holds, provided~$b^2$ is replaced by~$8$, 
and the maximal velocity in the strong field limit is 
\begin{align}
 \vm^\infty = \Vscale \sin(2\bet) \rondF(\lamhy/\luni),    \quad \mbox{for } b \gg 1.   
\end{align}
The velocity scale~$\Vscale$ introduced in Eq.~\eqref{eq:Vscale} 
is thus the maximal fluid velocity observable, 
that would occur for $\beta=\pi/4$ and a capillary of infinite thickness.

\begin{figure}[t!] 
\center
\includegraphics[width=0.42\textwidth]{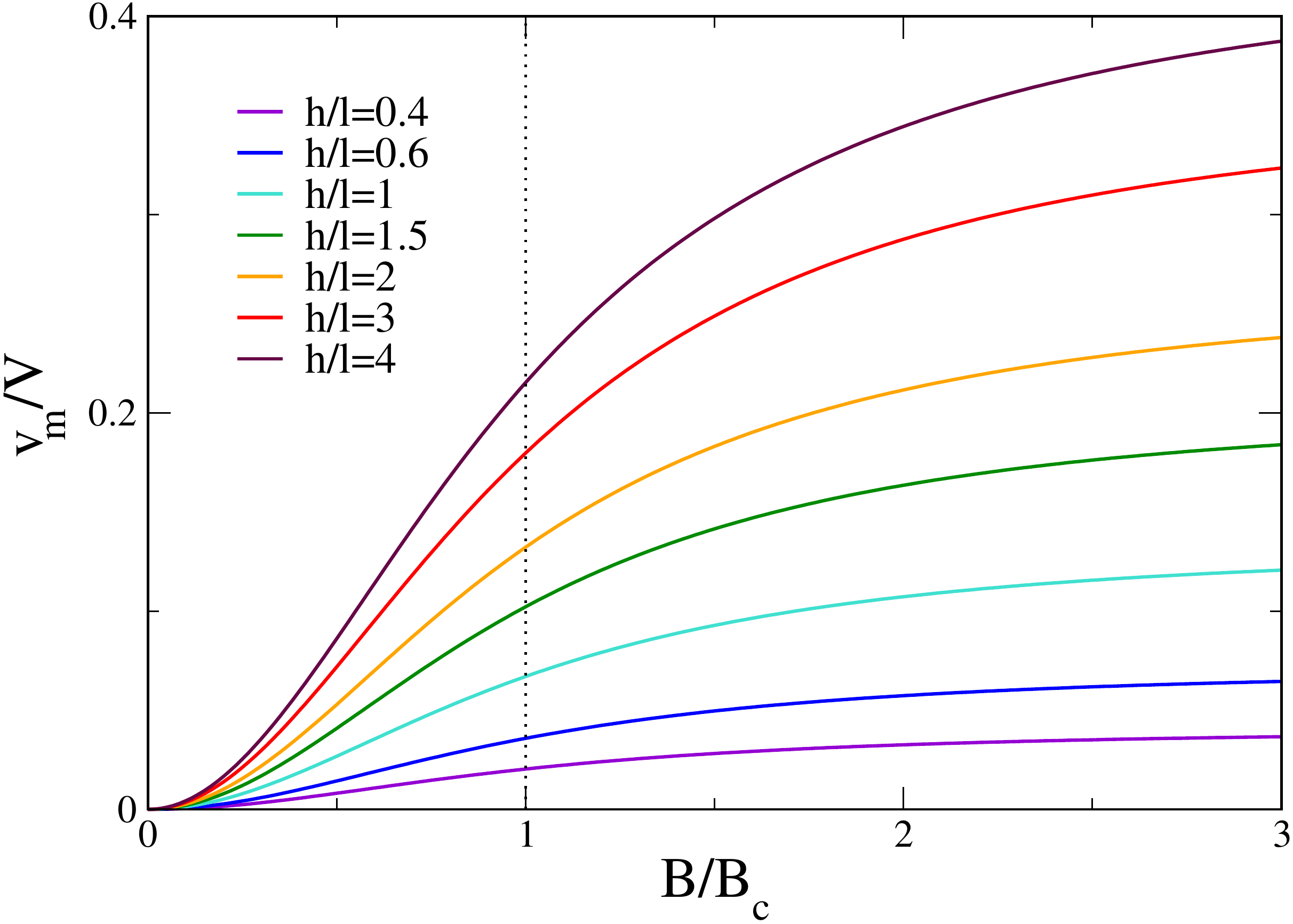}  
\caption{
\capti{Influence of magnetic field on flow magnitude.}
Maximal velocity~$\vm$ as a function of magnetic field 
for $\bet=\pi/4$ and various ratios between capillary half-thickness~$h$ and band width~$\len$. 
The velocity scale~$\Vscale$ is defined in Eq.~\eqref{eq:Vscale} 
and the characteristic magnetic field~$\Bchat$ is introduced in Eq.~\eqref{eq:Bchat}.  
}  
\label{fig:interpol}
\end{figure}

\subsubsection{Interpolation between weak and strong field}

With the weak and strong field limits in hand, 
and since the velocity profile keeps a similar form, 
one can  propose in the intermediate regime
a simple interpolation for the maximum fluid velocity, namely 
\begin{subequations}
\begin{align}
\vm  & =  \Vscale \sin(2\bet) \; \rondF(\lamhy/\len) \, \frac{\bh^2}{1  + \bh^2},   \label{eq:interpol} \\
\bh & \eqdef \frac{B}{\Bchat} \eqdef \frac{\mo B}{2^{3/2} \Gamo \lamo}.             \label{eq:Bchat}
\end{align}
\end{subequations}
The characteristic field $\Bchat$ delineates the boundary between the low and high field regimes. 
In particular, 
the velocity $\vm$ at $\Bchat$ is half the plateau value reached in the limit of strong field.   
Figure~\ref{fig:interpol}a illustrates the relation 
between the maximal fluid velocity and the magnetic field magnitude, 
taking  $\bet=\pi/4$ and  various capillary thicknesses.

\subsection{Comparison to experiments}

\subsubsection{Model limitations}

We are finally in a position to compare our results 
to the  experimental observations of Marmol \etal~\cite{Marmol_arxiv-2025}. 
As a preliminary to this discussion, 
let us point three limitations of our description.  

First, we assumed two-dimensional orientation of bacteria. 
In experiments, three-dimensional orientation is expected because shear along two directions ($x$ and $z$) 
induces bacteria rotation. 
Accounting for Jeffery orbits in this situation 
actually defines an already intricate problem~\cite{Vincenti_prf-2018,Talbot_pre-2024}, 
hence our choice to discard them entirely.   
Such an assumption  might be less of a problem when considering, as we do here, 
the velocity in the capillary midplane ($z=0$),  where  vertical shear is presumably low or vanishing. 

Second, the band studied here and in experiments 
are not identical in nature~\footnote{
One could call intrinsic 
those bands where the oxygen profile is dictated only by bacteria from the band, 
and extrinsic the bands where oxygen is also governed by external factors, 
such as imposed gradients in a microfluidic experiment 
or the presence of additional bacteria of different species, as may happen in natural habitats.}.
The former is a true steady state, where all bacteria participate to the band. 
The latter is a slowly evolving but transient state, where only a tiny fraction of bacteria 
are part of the band~\footnote{
From  Fig.~1b of Ref.~\cite{Marmol_arxiv-2025} and taking an experimental cell with length $L=8\,$mm, 
the fraction of bacteria in the band is less than $0.5$\%.}. 
The overwhelming majority of bacteria
is still distributed in their initial state  
as a homogeneous background that actually dictates the oxygen concentration.  
Nevertheless, with the toy sensing and binary sensitivity model used here, 
all that matters is whether the oxygen concentration is above or below its target value~$\cs$. 
In that particular case, it is still meaningful to compare the bands obtained in the two situations. 

Third, the experimental band has a weak but non-negligible asymmetry, which is visible 
in the bacteria density, fluid velocity and field direction dependence, 
all of which are not invariant upon field reversal. 
Such an asymmetry could only be accounted for by extending the present model, 
which is intrinsically symmetric. 

\subsubsection{Quantitative comparison and microscopic parameters}

A direct test of our prediction is not straightforward
because the microscopic properties of \Mgryphiwaldensefull{} 
were  not measured in the experiments of Ref.~\cite{Marmol_arxiv-2025}. 
We thus use the experimental observations 
to infer a possible set of bacterial parameters 
and show that they fall in the expected range of values. 
We discuss in turn each property.  

\paragraph*{(i) Density profile.}
If neglecting the asymmetry and discarding the background, 
the experimental bacterial profile is close to a Gaussian with standard deviation $15\,$\microm. 
The Laplace band can only provide an approximate description but remains qualitatively reasonable,  
in particular if some rounding occurs as discussed in Sec.~\ref{sec:continuouskernel}. 
The band width $\len = \vo \lamoi/\alb \ko = 15\,$\microm{} 
would be compatible with 
$\vo=15\,$\micromsi~\footnote{
The distribution of velocity may in fact be bimodal, 
as reported in Fig.~3 of Ref.~\cite{Reufer_bpj-2014} for South Seeker MTB, 
but not necessarily for North Seeker MTB.}, 
$\lamoi = 1\,$\si, 
$\ko=1/2$ and $\al=-1$, i.e perfect reversals.  
The total number of bacteria in the band, expressed per unit area of lateral  surface, 
is in the range $\Ntotlat = 0.5-5\,10^{11}\,$m$^{-2}$. 
Note that according to Eq.~\eqref{eq:vfa}, 
the uniform background of bacteria makes no contribution to  the fluid flow. 

\begin{figure}[t!] 
\center
\includegraphics[width=0.42\textwidth]{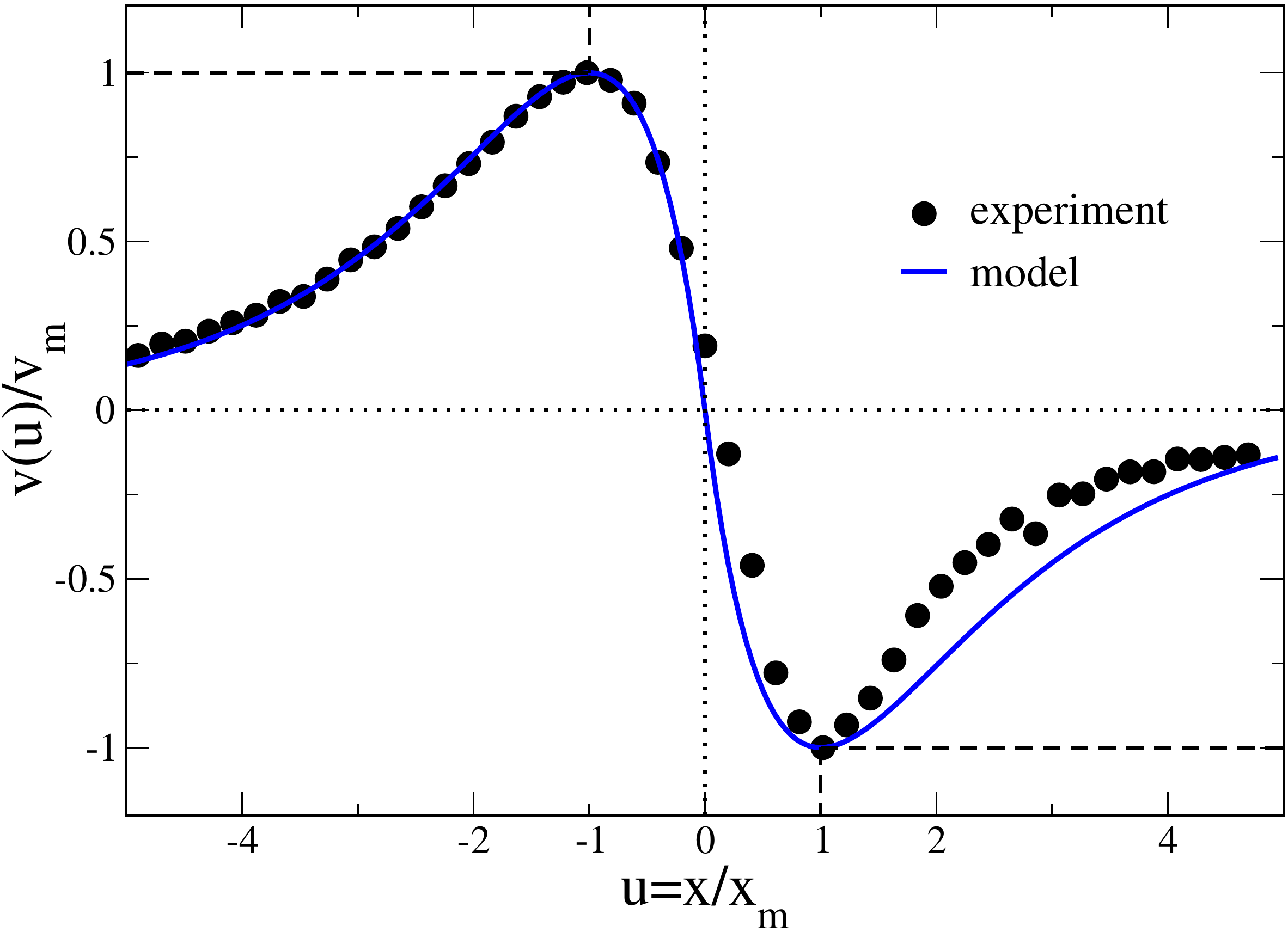}  
\caption{
\capti{Comparison of flow velocity  in theory and experiments}. 
When scaled by the maximum position~$\xm$, which varies by about 10\%,  
and by the maximum velocity~$\vm$, which varies over almost a decade,
velocity profiles measured at various bacteria densities and magnetic fields  
approximately collapse on each other~\cite{Marmol_arxiv-2025}.  
Points show the resulting master curve,   
obtained as an average over many experimental profiles~\cite{Ybert_personnalcommunication-2025}.
The line is the prediction derived from Eq.~\eqref{eq:vx}, without any free free parameter. 
}  
\label{fig:expvelocityprofile}
\end{figure}

\paragraph*{(ii)~Velocity profile.}
When rescaled by its maximum value, 
the fluid velocity profile 
is predicted to depend  on two parameters only: 
the band width~$l$ and 
the hydrodynamic length~$\lamhy$ fixed by the capillary thickness $2h=95\,$\microm.  
This expectation is borne out in experimental data~\cite{Marmol_arxiv-2025} 
and allows to define a master curve, 
which is shown with points in Fig.~\ref{fig:expvelocityprofile},  
together with the theoretical prediction derived from Eq.~\eqref{eq:vx}.  
While deviation on the low-oxygen side of the band can reach 30\% , 
the agreement on the high-oxygen side is excellent. 
The overall reasonable agreement, without any fitting parameter, 
suggests that the hydrodynamic treatment of the problem is adequate~\footnote{
Because the hydrodynamic length scale $\lamhy$ is significantly larger than the band, 
the flow profile is not very sensitive to the precise shape of the density profile, 
as long as it can be characterized by a single length scale~$l$.}. 

\paragraph*{(iii) Dependence on field orientation}
Except for the slight asymmetry, the experimental flow profile varies as $\sin(2\beta)$, in agreement with prediction. 
\paragraph*{(iv) Dependence on bacteria number.}
The Fokker-Planck equation used throughout this work is linear  with respect to the bacteria distribution 
and the flow is directly proportional to the forcing. 
Our approach is thus limited to the regime linear in bacteria number, 
which is observed in experiments up to $\Ntotlat$  around $2.5\,10^{11}\,$m$^{-2}$.

\paragraph*{(v) Maximal velocity.}
The linear dependence between bacteria number and maximal velocity 
allows to estimate the force dipole moment~$\sigo$. 
Exploiting Eq.~\eqref{eq:interpol}, where there is no other unknown parameter~\footnote{
We consider the limit of strong field $\bh \gg 1$, which is closer to the experimental data available. 
}, 
one obtains 
$\sigo = -1.5\,10^{-19}\,$J~\footnote{
Apart from the basic modelling assumptions, such as the type of reorientation event,   
the difference with the higher value inferred in Ref.~\cite{Marmol_arxiv-2025} can have two origins: 
a distinct treatment of the strong field limit 
and a more accurate hydrodynamic Green function, which does not rely on a lubrication approximation. 
Note that both  Ref.~\cite{Marmol_arxiv-2025} and the present work 
assume bacterial orientation that are two-dimensional. 
}, 
which has the correct order of magnitude~\footnote{
On dimensional grounds~\cite{Saintillan_expmech-2010}, 
the force dipole can be written as $\sigo = A \eta \vo a^2$, 
with $a$ a characteristic size and $A$ a numerical constant, 
unknown in general  and likely to be species-dependent. 
For  an {\it Escherichia coli} bacteria swimming at average velocity $\vo = 22\,$\micromsi, 
Drescher \etal~\cite{Drescher_pnas-2011} found a force $F \simeq 0.4\,$pN and size $d \simeq 2\,$\microm, 
giving a moment $\sigo = - F d \simeq  - 8\,10^{-19}\,$J. 
For a swimming velocity $\vo = 15\,$\micromsi, one would thus expect a moment $\sigo$ around  $-5\,10^{-19}\,$J. 
Such an estimate is valid for the multiflagellated {\it E. coli} 
but may not apply to the biflagellated \Mgryphiwaldense.
}, 
and as can be expected from our series of approximations, 
lies in the lower side of the admissible range~\footnote{
There are several reasons to believe  that our estimate for $\sigo$ is a lower bound.  
First, 
whereas the model considers the fluid velocity in the midplane $z=0$, where it is presumably the highest, 
the experimental velocity is averaged over a finite thickness $\Delta z$, 
which may not be negligible with respect to capillary thickness. 
The reported velocity may thus be underestimated with respect to the maximum.  
Second, 
our simplifying assumptions have discarded a number of couplings 
that would counteract the flow.  
In particular, the non-spherical shape of \Mgryphiwaldense{} 
may induce rotation by the flow that is stronger than predicted here. 
Third, 
the model neglects lateral confinement in the $y$-direction,  
and any imperfection such  as non-straight bands, 
which would presumably reduce the flow.  
We conclude that the magnitude of the dipole force moment 
is at least $|\sigo| =  1.5\,10^{-19}\,$J 
and is likely to be higher.
}. 
Note that the fixed value of dipole moment, assumed throughout the model, 
is permitted by the biflagellated nature of \Mgryphiwaldense{} 
and that the negative $\sigo$ is indicative of a pusher swimmer.

\paragraph*{(vi) Characteristic field.}  
As defined in Eqs.~\eqref{eq:interpol}-\eqref{eq:Bchat}, 
the characteristic field $\Bchat$ is where the fluid velocity reaches half its strong-field plateau value.  
Taking a magnetic dipole 
$\mo=2\,10^{-16}\,$J$\,$T\inv~\footnote{
Depending on the method employed, 
the estimate of the magnetic moment can cover almost a decade, 
as reported by Ref.~\cite{Nadkarni_plosone-2013} on the bacteria {\it Magnetospirillum magneticum}. 
We used the value of Ref.~\cite{Reufer_bpj-2014} for \Mgryphiwaldense{}, 
see also~Ref.~\cite{Pichel_jmmm-2018}.   
}, 
and a rotational friction coefficient~$\Gamo=7\,10^{-20}\,$J$\,$s~\footnote{
For a sphere of radius $\as=1.4\,$\microm, 
$\Gamo=8 \pi \eta \as^3 \simeq 7\,10^{-20}\,$J$\,$s. 
For an ellipsoid  with semi-axis lengthes~$\arod=5\,$\microm{} and ~$\brod=1\,$\microm, 
$\Gamo= \pi \eta \arod^3/(3 \log(2\arod/\brod)-1/2)  \simeq 7\,10^{-20}\,$J$\,$s~\cite{book_Berg-RandomWalksBiol}. 
},  
one finds  $\Bchat \simeq 1\,$mT, compatible with experimental values. 

To conclude, our model, though analytical, is  consistent 
with experimental observations with a choice of microscopic parameters that appears reasonable, 
suggesting that the description chosen is correct, on a semi-quantitative level at least. 

\section{Summary and perspectives}
\label{sec:conclusion}

To summarize, 
we developed a Fokker-Planck description of the steady aerotactic band of swimming bacteria.  
The full problem  is an intricate one, as it couples multiple physical phenomena,  
including  oxygen consumption and diffusion, 
self-propulsion and tumbling  with rate modulated by aerotaxis. 
In addition, magnetotactic bacteria also involve orientation along the magnetic field 
and possibly advection and rotation by the induced fluid flow. 
Throughout our treatment, we made a number of simplifying assumptions 
so as to partially decouple the phenomena and keep analytical tractability. 

Our main result is the Laplace model of aerotactic band, 
an exact solution to the Fokker-Planck equation, obtained for aerotactic binary sensitivity. 
Such simple behavior encapsulates the preference for a target oxygen concentration commonly observed in bacteria.   
Though not applicable to all bands, 
the Laplace band  could serve as a convenient model to investigate, at the qualitative level,  
the generic properties of aerotactic bands.  
We made a first step in this direction in the present work 
by investigating the aerotactic band of magnetotactic bacteria.  
Treating the magnetic field as a small perturbation,  
we obtained an extremely simple prediction for the fluid flow.  
Our predictions are consistent with many features reported in experiments.  

Given the physical ingredients involved, 
it is worth pointing out the peculiar nature of fluid flow in aerotactic bands of magnetotactic bacteria.  
In contrast to collective behaviors previously observed with magnetotactic bacteria, 
such as swarms~\cite{Belovs_pre-2017,Birjukovs_arxiv-2024}, dynamical patterns~\cite{Koessel_njp-2020} 
and clusters or plumes~\cite{Guell_jtb-1988,Thery_sr-2020,Pierce_prl-2018}, 
self-organization here is not driven by hydrodynamic interactions. 
Instead, self-assembly originates from a different mechanism 
-- aerotaxis and oxygen consumption -- and 
fluid flow emerges as a by-product of the self-assembled structure.  

This work opens a number of perspectives for future investigations.   
A first direction is to extend the model by relaxing some of its basic assumptions. 
For instance, we assumed everywhere a single velocity whereas 
distinct forward and backward swimming modes 
could result in different velocities~\cite{Reufer_bpj-2014}. 
Likewise, 
we considered swimmers with constant force dipole, 
but some swimming strategies, such as run-reverse for monoflagellate bacteria,  
alternate between pusher and puller modes~\cite{Son_natrevmb-2015}.    
An other direction to pursue is  the quantitative description of real bands 
by inferring from experimental data the aerotactic kernel. 
Given the differences in band profiles already observed~\cite{Lefevre_bpj-2014}, 
one can anticipate some significant variation between species. 
Besides, we focused exclusively on bands in steady state,  
which is more amenable to a complete experimental characterization, 
but the transient regime during which the band emerges and moves    
also deserves some investigation. 

Finally, 
bands of magnetotactic bacteria  still raise peculiar challenges.  
The regime of intermediate magnetic field, which is most relevant experimentally, remains to be addressed.  
A full description accounting for three-dimensional orientation of bacteria and non-spherical shape  
is also needed. 
The most likely route to do so is to resort to numerical resolution of the Fokker-Planck equation.  
Such an endeavor, that is not straightforward  and may require specialized techniques, 
is left for a dedicated study.

We conclude with a speculation on the biological usefulness 
of fluid flow in aerotactic bands of magnetotactic bacteria. 
Given the magnitude of Earth magnetic field, 
naturally occurring bands would induce a flow velocity roughly 100 times slower than in experiments, 
thus on the order of $0.1\,$mm\,\si, with  higher values possible in denser bands. 
Such an advection drift, though weak,   could induce convection cells 
and could  dominate diffusive transport over large length  and time scales. 
Would the resulting  fluid transport  be beneficial to  bacteria 
by providing, for instance, a steady current of nutrients? 
Besides enhanced aerotaxis~\cite{Lefevre_mmbr-2013} and 
navigation in porous environment~\cite{Codutti_elife-2024}, 
could the emergence of band flows be the key evolutionary advantage 
that has driven bacteria to possess an inner magnet? 
If so, 
aerotactic bands would not only represent a self-organized structure of bacteria alone 
but also a way to tailor their  environment and ecological niche. 

\vspace*{3mm}

\paragraph*{Acknowledgments.} 
I am grateful  to Malo Marmol, Damien Faivre and Cécile Cottin-Bizonne   
for early access to the experimental results of Ref.~\cite{Marmol_arxiv-2025}. 
Many thanks are due to Christophe Ybert for inspiring discussions, 
for providing the data of Fig.~\ref{fig:expvelocityprofile}, 
and for sharing many insights on the experiments. 
Exchanges with \'Eric Clément are also  gratefully acknowledged. 
Financial support includes  ANR-20-CE30-0034 BACMAG and ETN-PHYMOT within the European Union’s Horizon 2020 research and innovation programme under the Marie Sk\l odowska-Curie grant agreement No 955910. 

\appendix 

\section{Governing equation for one-dimensional motion}
\label{sec:app:1D}

We derive  the governing equation for 
bacteria that move only perpendicular to the band and reverse direction at each tumble. 
This model was used previously in Refs.~\cite{Smith_bpj-2006,Bennet_plosone-2014,Lefevre_bpj-2014} 
and would be relevant for a run-reverse bacteria whose direction of motion is imposed 
by a sufficiently strong magnetic field. 
The density probabilities of leftward and rightward-moving bacteria, $L(x,t)$ and $R(x,t)$ respectively, obey: 
\begin{subequations}
\begin{align}
\partial_t R + \vo \partial_x R &= - \lamrl R +  \lamlr L,  \label{eq:1DFPa}   \\
\partial_t L + \vo \partial_x L &= + \lamrl R -  \lamlr L,  \label{eq:1DFPb}
\end{align}
\end{subequations}
where is $\lamrl$ the transition rate from right-going to left-going motion. 
Introducing the total bacterial density $p=R+L$ and $J=R-L$ leads to
\begin{subequations}
\begin{align}
\partial_t p + \vo \partial_x J    &= 0,                             \\
\partial_t j    + \vo \partial_x p &= p (\lamlr-\lamrl)- J(\lamlr+\lamrl).  
\end{align}
\end{subequations}
Seeking a steady state and noting that the bacterial current vanishes ($J=0$) in a closed capillary, 
one finds that the density $p(x)$ satisfies 
\begin{align}
\vo \frac{p'}{p}   &=\nu(x) \equiv \lamlr(x)-\lamrl(x).     
\end{align}
Within our model, the spatial modulation in reverse rate is $\nu(x) = \lamo \krn(x)$,  
which gives back Eq.~\eqref{eq:pxtheta} with $\alb=1$.

\section{Toy sensing approximation}
\label{sec:app:toysensing}

Here we discuss simple forms of the aerotactic response 
and the approximation leading to toy sensing.
Let us  consider a microorganism that can detect both the local concentration $c(x)$ 
and the gradient $c'(x) \costheta$ along the body or velocity direction~\footnote{
It is assumed, as everywhere else, that the concentration varies only along the $x$-axis.}. 
The aerotactic behavior, when assumed local and instantaneous, 
is defined by a response function $\Res(c, c' \costheta)$. 
For simplicity, $\Res$ is taken as a separable function of its arguments, 
that is $\Res_0(c) \Res_1(c' \cos\theta)$. 
As regards~$\Res_0$, the simplest choice is binary sensitivity $\Res_0(c) \sim \sign(\cs - c)$. 
As regards~$\Res_1$, there are two simple behaviors to consider:  

\paragraph*{(i) Linear response.} 
In the limit of small gradients, a linear dependence can be expected, 
with $\Res_1 \sim c' \costheta$. 
Such a behavior corresponds to the linear sensing investigated in Sec.~\ref{sec:linearsensing}. \\

\paragraph*{(ii) Saturation.}
In the limit of large gradient, the response could saturate, 
with  $\Res_1 \sim  \sign( \costheta)$, up to a constant numerical prefactor. 
Such a form appears more difficult to handle analytically. 
Therefore, it is convenient to approximate $\sign( \costheta)$ by $\costheta$, 
which amounts to retain only the first term of the Fourier series. 
The resulting behavior  $\Res_1 \sim  \costheta$ gives the toy sensing model 
which is considered in Sec.~\ref{sec:toysensing}. 
While only an approximation of the real aerotactic behavior~\footnote{
The toy sensing approximation implies that the microorganism 
can detect gradient direction, 
a feature that is not expected if aerotaxis relies on temporal sensing 
but is conceivable for large microorganisms endowed with  spatial sensing 
if they can sample concentration differences in various directions. }, 
it has the advantage of tractability and leads to the simplest band structure.

\begin{widetext}

\section{Derivation of the low-field expansion}
\label{sec:app:lowfield}

We consider the Fokker-Planck equation 
\begin{align}
 \partial_t p =& - \vo \costheta \, \partial_x p  + \difrot \partial^2_{\theta\theta} p  
                 + b \, \partial_\theta \left[   \sin(\theta-\bet) p \right]  
                 - \lam(x,\theta)  p + [ \lam(x,\theta)  p] \circ  h,    \label{eq:FPfield}
\end{align}
and for low magnetic field~$b$, we seek an expansion of the form 
\begin{align}
 p(x,\theta) = \sum_{m=0}^\infty p_m(x,\theta) b^m,    \label{eq:bexpansion}
\end{align}
Working with $x>0$ for simplicity,  and gathering terms by order, one gets  
\begin{align}
 \left[ \costheta \partial_x  + (1 + \kom \costheta)  - \difrot \partial^2_{\theta\theta}  \right] p_m
 - \frac{\intI_m(x)}{2\pi} 
   =  \partial_\theta \left[  p_{m-1} \sin(\theta-\bet ) \right],  
 \quad 
 \intI_m(x) \eqdef \intpi (1 + \kom \costheta') p_m(x,\theta') \dd \theta'.     \label{eq:bexpansion-rec}
\end{align}
The solving is recursive: if $p_{m-1}$ is known, 
the next order term $p_m$ is obtained by solving the integro-differential equation Eq.~\eqref{eq:bexpansion-rec}. 
To do so, 
we introduce two types of transform. 
For position~$x$, we take the Laplace transform ($x \to \xil$). 
A function $f(x)$ is related to its  Laplace transform  $f(\xil)$ through 
\begin{align}
f(\xil)      = \int_0^\infty   e^{-\xil x} f(x) \, \dd x ,     \qquad \quad 
f(x)        = \frac{1}{2 \pi i} \int_{c-i\infty}^{c+i\infty}    e^{x \xil} f(\xil) \,\dd \xil. 
\end{align}
For orientation angle~$\theta$, we use the Fourier series representation, defined from 
\begin{align}
 f(l)      = \frac{1}{2\pi} \int_{-\pi}^{\pi}   e^{-i l \theta} f(\theta)   \, \dd \theta,   \qquad \quad 
f(\theta) =  \sum_{l=-\infty}^{\infty} f(l) \: e^{i l \theta}. 
\end{align}
Note that for both types of transforms, 
the name of the variable indicates the nature of the function. 
When switching from $p(x,\theta)$ to $p(\xil,l)$ function, 
Eq.~\eqref{eq:bexpansion-rec} yields 
\begin{align}
 (1 + \difrot) f_m(l) + (\xil+\kom) \opT f_m(l)  = \opT \fmid_m(l) + \left[  f_m(0) + \kom \opT f_m(0) \right] \delta_l 
 + \frac{l}{2} \left[  e^{-i\bet} S(l-1) -  e^{i\bet} S(l+1)  \right],   \label{eq:bexpansion-l}
\end{align}
where we introduced the short-hand notations 
\begin{align}
 f_m(l) \eqdef p_m(\xil,l),  \qquad  \fmid_m(l) \eqdef p_m(x=0,l),  \qquad S(l) \eqdef p_{m-1}(l,\xil),  
 \qquad 
 \opT f(l) \eqdef \frac{f(l+1)+f(l-1)}{2}.  
\end{align}
$\fmid_m(l)$ describes the orientation distribution in the band center,  
$S(l)$ is the source term and $\opT$ is a shifting operator on the $l$~variable. 

The first step to solve Eq.~\eqref{eq:bexpansion-l} is to postulate  
that the Fourier expansion has a finite number of terms. 
Specifically, 
the $m$-order term $p_m(\xil,l)$ has non-vanishing Fourier coefficients only for $|l| \leqslant m$. 
With this assumption, one can solve the system  of equation 
and express  $f_m(l)$ a function of $\fmid_m(l)$.  
Now, the choice of $\fmid_m(l)$ must obey several constraints, 
the first of which are symmetry relations. 
The distribution $\fmid_m(\theta) \eqdef p_m(x=0,\theta)$ must be invariant 
in the transformation $b \to -b$ and $\theta \to \theta + \pi$~\footnote{
Because it simply switches the left and right side of the band.}, 
which from Eq.~\eqref{eq:bexpansion}, gives $\fmid_m(\theta) = (-1)^m \fmid_m(\theta+\pi)$.   
This  implies that Fourier components $\fmid_m(l)$ with $m$ even (resp. odd) 
vanishes for $l$ odd (resp. even). 
In addition to the symmetry requirement, 
$\fmid_m(l)$ must be normalizable, 
implying that the bacterial density must vanish far from the band ($x \to \infty$). 
A positive pole $\xilp$ in the rationale function $p_m(\xil,l)$
would lead in $p_m(x,l)$ to a term of the form $\exp(\xilp x)$, 
which is not physically acceptable. 
We thus choose $\fmid_m$ so that  $p_m(\xil,l)$ has only strictly negative poles. 
Finally, when considering the band on both sides, the probability density must be normalized. 

Taken together, 
those requirements are sufficient to determine for $p_m(\xil,l)$ a unique solution at order $m=1$ and $2$. 
After coming back to original $(x, \theta)$ variables, the end result is 
\begin{subequations}
\begin{align}
p_1(x,\theta) &=  \frac{\kom e^{-\kom x} }{4 \pi  (1+\difrot)} \left[ \sin \bet  \sin \theta  + (1+\difrot) x \cos \bet  \right], \\
p_2(x,\theta) &=  \frac{\kom e^{-\kom x}}{8 \pi    \difrot^{[1,4]} } 
\left[ \difrot^{[1,4]} \left( x^2-\frac{2}{\kom^2}\right) \cos ^2\bet  + \difrot^{[4]} x \sin (2 \bet ) \sin \theta -2 \sin ^2\bet  \cos (2 \theta ) +  \sin (2\bet) \sin (2 \theta)/2 \right], 
\end{align}
\end{subequations}
with $\difrot^{[1,4]} \eqdef  (1+\difrot) (1+4\difrot)$ and   $\difrot^{[4]} \eqdef 1+4 \difrot$.  
Equations~\eqref{eq:pxthetafieldA}-\eqref{eq:pxthetafieldC} are equivalent expressions 
at second order in field $b$, where rotational diffusion is neglected ($\difrot=0$).

\section{Exact expression for hydrodynamic Green function $G(x)$}
\label{sec:app:G}

The Green function as defined from Eq.~\eqref{eq:eqG} for the flow velocity 
in the midplane of the capillary is 
\begin{align}
G(x) = \frac{h}{\pi^2} \left\{ 
 |\xt| \left(-\pi + 4 \arctan \left[ e^{-|\xt|} \right] + 4 \arctan \left[  \tanh(|\xt|/2) \right]    \right)  +   2 i \Li_2[-i e^{-|\xt|} ] - 2 i \Li_2[i e^{-|\xt|} ] 
\right\},   
\quad 
\xt \eqdef \frac{\pi}{2h} x,               \label{eq:Gexact}
 \end{align}
where $\Li_m$ is the polylogarithm of order~$m$. 
To obtain this expression,  
one can first consider $-q^2 G(q)$, whose inverse Fourier transform can be computed using formula 3.981.3 of Ref.~\cite{book_GradshteynRyzhik-TableInt} and yields $G''(x)$. 
Integrating twice while enforcing boundary conditions and parity leads to Eq.~\eqref{eq:Gexact}. 

\end{widetext}

\end{document}